\documentclass[sigconf]{acmart}

\usepackage{xspace}
\usepackage{listings}
\usepackage{enumitem}
\AtBeginDocument{%
  \providecommand\BibTeX{{%
    \normalfont B\kern-0.5em{\scshape i\kern-0.25em b}\kern-0.8em\TeX}}}

\setcopyright{acmcopyright}
\copyrightyear{2024}
\acmYear{2024}
\setcopyright{rightsretained}
\acmConference[CHI '24]{Proceedings of the CHI Conference on Human Factors in Computing Systems}{May 11--16, 2024}{Honolulu, HI, USA}
\acmBooktitle{Proceedings of the CHI Conference on Human Factors in Computing Systems (CHI '24), May 11--16, 2024, Honolulu, HI, USA}
\acmDOI{10.1145/3613904.3642172}
\acmISBN{979-8-4007-0330-0/24/05}

\begin{document}

\title{Epigraphics: Message-Driven Infographics Authoring}

\author{Tongyu Zhou} 
\authornote{Work done during internship at Adobe Research.}
\affiliation{%
 \institution{Brown University}
 \city{Providence}
 \state{Rhode Island}
 \country{USA}}
 \email{tongyu_zhou@brown.edu}

\author{Jeff Huang}
\affiliation{%
  \institution{Brown University}
  \city{Providence}
  \state{Rhode Island}
  \country{USA}}
  \email{jeff_huang@brown.edu}

\author{Gromit Yeuk-Yin Chan}
\affiliation{%
  \institution{Adobe Research}
  \city{San Jose}
  \state{California}
  \country{USA}}
  \email{ychan@adobe.com}

\definecolor{alizarin}{rgb}{0.82, 0.1, 0.26}
\newcommand{\punchline}[1]{\textcolor{alizarin}{\underline{#1}}\newline}

\definecolor{lightpink}{RGB}{237,157,202}
\definecolor{lightred}{RGB}{210,121,121}
\definecolor{lightorange}{RGB}{230,170,50}
\definecolor{lightgold}{RGB}{210,194,121}
\definecolor{lightgreen}{RGB}{121,210,121}
\definecolor{lightaqua}{RGB}{121,206,210}
\definecolor{lightblue}{RGB}{121,124,210}
\definecolor{lightpurple}{RGB}{153,102,255}
\definecolor{red}{RGB}{178,34,34}
\definecolor{gray}{RGB}{166,166,166}

\newcommand{\todo}[1]{\textcolor{red}{[TODO] \emph{#1}}}
\newcommand{\systemname}{\textsc{Epigraphics}\xspace}
\newcommand{\gromit}[1]{\textcolor{lightblue}{[Gromit] \emph{#1}}}
\newcommand{\tongyu}[1]{\textcolor{lightpink}{[Tongyu] \emph{#1}}}
\newcommand{\chirevision}[1]{#1}
\renewcommand{\shortauthors}{Zhou et al.}

\begin{abstract}
  The message a designer wants to convey plays a pivotal role in directing the design of an infographic, yet most authoring workflows start with creating the visualizations or graphics first without gauging whether they fit the message. To address this gap, we propose \systemname, a web-based authoring system that treats an ``epigraph'' as the first-class object, and uses it to guide infographic asset creation, editing, and syncing. The system uses the text-based message to recommend visualizations, graphics, data filters, color palettes, and animations. It further supports between-asset interactions and fine-tuning such as recoloring, highlighting, and animation syncing that enhance the aesthetic cohesiveness of the assets. A gallery and case studies show that our system can produce infographics inspired by existing popular ones, and a task-based usability study with 10 designers show that a text-sourced workflow can standardize content, empower users to think more about the big picture, and facilitate rapid prototyping.
\end{abstract}



\begin{teaserfigure}
  \includegraphics[width=\textwidth]{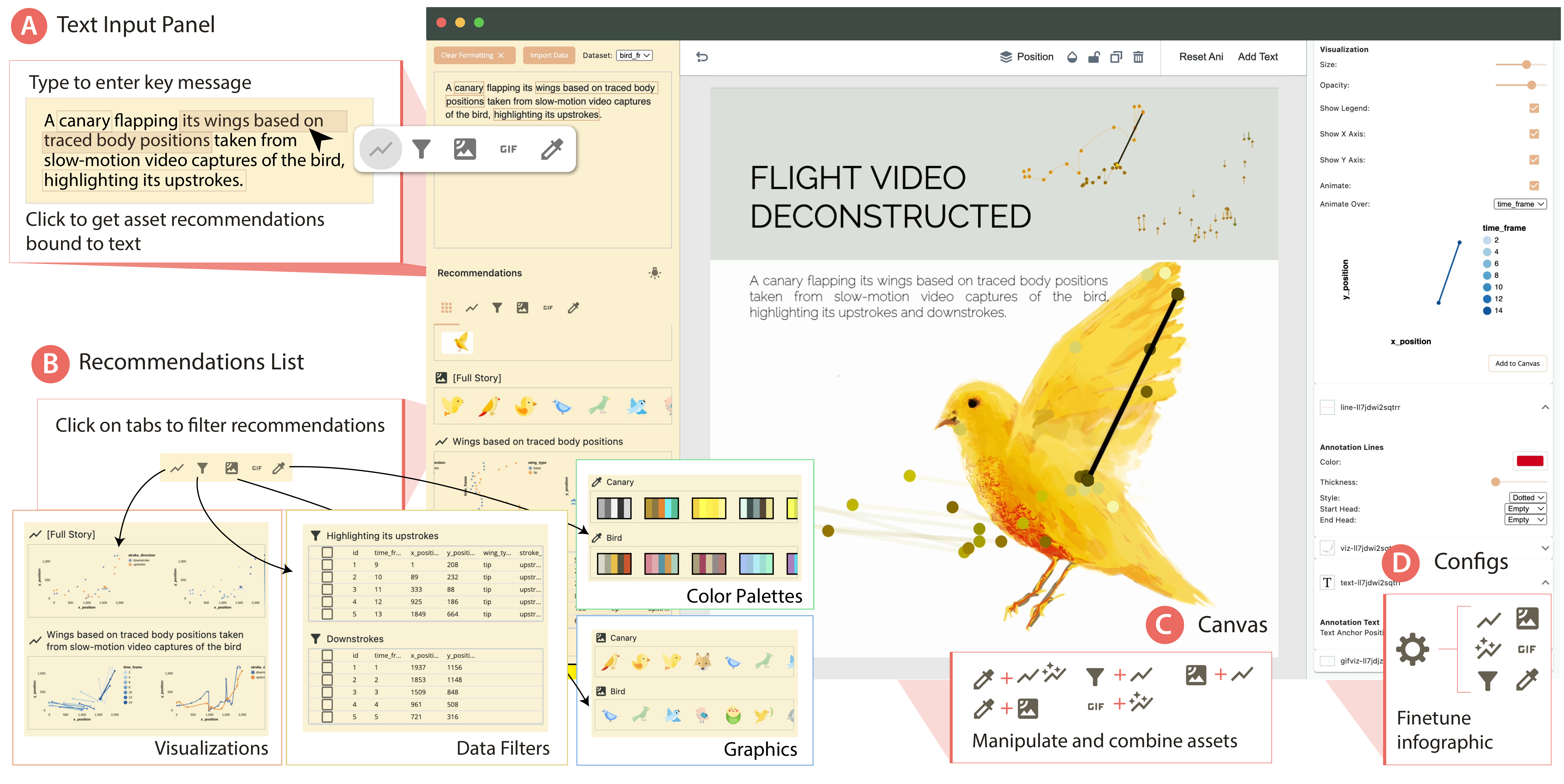}
  \caption{To create an infographic with \systemname, the user types a key message they wish to convey (A), automatically receives asset recommendations based on brushed text chunks (B), adds desired assets to the canvas to manipulate and merge (C), and finally fine-tunes the assets until they are satisfied (D).}
  \label{fig:user-interface}
  \Description{Teaser figure showcasing the user interface of Epigraphics. The left-hand panel displays a text input box and a list of generated recommendations. The middle is a canvas containing a completed infographic of a canary’s wingspans through flight. The right-hand panel is a list of existing assets on the canvas with toggles, dropdowns, and sliders to change the configurations of each asset. From this interface, there are 4 popups further explaining what the text input box, recommendations list, canvas, and configuration panels do.}
\end{teaserfigure}

\maketitle

\section{Introduction}


The infographic is an evocative, visual vignette of data. By synthesizing visualizations, illustrations, images, color, and text, it condenses datasets into digestible takeaways and stories that can be readily consumed by viewers. Unlike other related mediums such as data videos \cite{amini2017authoring}, infographics usually deliver a singular message to an audience who may not necessarily have the time nor background to analyze and draw their own conclusions about data. By minimizing the cognitive load required to interpret information, they motivate viewers to engage in deeper reflections about information \cite{clark2010graphics} and have seen usage beyond data science and design into tangential fields such as contemporary journalism \cite{fu2023more} and education \cite{chuda2007visualization, firat2018towards}.



However, designing infographics can be difficult, time-consuming, and unintuitive. Current workflows involve creating, arranging, and editing visualizations, graphics, and text components based on the message the designer wants to convey. Although this message is the main mechanism that drives the authoring process, most workflows start with the drawing \cite{xia2018dataink} or visualization \cite{wang2018infonice} first. Then, designers may fill in complementary titles, captions, or annotations afterwards to draw relationships between visual elements. This is oftentimes an iterative process where elements such as charts, illustrations, text, and overall design are created, passed on for peer feedback~\cite{cambre2018juxtapeer, kulkarni2015peerstudio, bawabe2021uxfactor}, then recreated multiple times to complement the message and strike a balance between information and aesthetics.

But what if the message starts and remains as a constant focus in the entire infographic design process?
Authoring can then be more akin to storytelling, where the design is directly driven by the message the designer intends to convey. Given the increasing popularity of authoring systems for automated design in both research~\cite{guo2021vinci,tyagi2022infographics} and commercial products (e.g., Adobe Express\footnote{\url{https://www.adobe.com/express/}}, Figma\footnote{\url{https://www.figma.com/}}, Canva\footnote{\url{https://www.canva.com/}}), we believe a \textbf{text-based message} can be a powerful, interactive medium to \textbf{automate the creation of infographic components}. It can act as an anchor to retrieve appropriate features and visual representations of the data, as well as induce design themes, graphics, and highlights in the infographic. When thinking about the content and placement of the text first, 
the designer can transform their implicit design thinking and discovery into explicit design edits.
We note that in this context, we specifically mean a message that directly conveys meaning or themes of the infographic, rather than just a text-based prompt. Focusing on this message is analogous to writing out ``alt text'' first; by conceding some creative liberty to generative models and allowing them to fill in the gaps of the assets, the designer can create a more cohesive data story.

While there are existing tools that offer text as a starting point for crafting infographics \cite{cui2020text, qian2021retrieve, tyagi2022infographics} and strategies for constructing natural language and visualization couplings \cite{srinivasan2021collecting}, they all rely on existing infographic or visualization exemplars created by a third party. Visualization creation is thus a retrieve-then-modify task rather than a generative process. The resulting infographics produced are derivative, meaning the designer has reduced flexibility over the originality of their creations. 
Thus, we propose \systemname, a web-application where users can craft text-based key messages and brush over their constituents as sources to generate infographic assets such as chart primitives, graphics, color themes, data filters, and animations. They serve as recommendations for users to add to and rearrange on a canvas. Unlike other data agnostic design tools like PowerPoint or Figma, the text is always processed in the context of an imported dataset. The generated assets are modular--this is a deliberate design decision to 1) provide flexibility to pair-wise merge them and 2) preserve creative autonomy by not providing the entire design. The components can be merged via between-asset interactions such as recoloring, highlighting, animation syncing, and injecting graphics into visualizations as glyphs.
Using the system, the user can quickly produce visual elements aligned with \textit{both} the message and the data, and thus focus on the composition and symbolism of their design rather than on individual element aesthetics. We assess the efficacy of the system through \chirevision{a gallery with case studies} and a task-based usability study. The demonstrations illustrate the potential expressiveness and capabilities of the system in the hands of an experienced user, while the study reveals insights into workflow and first-time usage outcomes of novice users. Together, they reveal that a message-first authoring workflow for infographics is effective at standardizing content, does not provide a cohesive layout but promotes holistic thinking, and empowers rapid prototyping.
The key contributions of this paper include:
\begin{enumerate}
    \item A system that generates infographic assets based on a key message whenever the user interacts with text and supports within and between-component interactions.
    \item Design lessons extrapolated from case studies and a user study about how a message-sourced workflow can influence design processes and outcomes.
\end{enumerate}
\section{Related Work}

\subsection{Data to Beautiful Graphics}

Many systems that aid users in creating the beautiful graphics commonly found in infographics focus on binding raw data to visual representations. For example, Data-Driven Guides~\cite{kim2017datadriven} enable designers to create guidelines to which custom shapes can be linked. These guides can then encode data into each shape's length, area, and position and dynamically deform the drawing based on changing data. DataInk~\cite{xia2018dataink} adopts a similar approach of binding raw data to shapes, but instead of properties, the entire shape is treated as a singular glyph. Users can use direct manipulation to link free-form sketched glyphs to data points, which can then be used to author creative visualizations. DataQuilt~\cite{zhang2020dataquilt} further expands upon this by allowing users to create these bindings for real images such as paintings and photographs so that visualizations can adapt the appearance of collages. However, one limitation of such strategies is that the mappings themselves need to be recreated each time for new datasets. To resolve this, Charticulator~\cite{ren2019charticulator} additionally enables the export of mapped visualizations to be reused as templates with other data.

In instances where the dataset is convoluted, animations can effectively guide the viewer's gaze from one piece of information to another. Gemini~\cite{kim2021gemini} is one such recommendation system that allows users to write declarative grammar to produce animated transitions between related statistical graphs. Data Animator~\cite{thompson2021dataanimator} removes the need for coding altogether by automatically generating transitions between two static visualizations by matching objects and supports a visual interface for fine-tuning. Animated Vega-Lite~\cite{zong2023animated} alternatively reframes animated visualizations as time-varying data queries, where the time encodings map data fields to key frames. Outside of transitions, animations in graphs and visualizations can also be decorative to draw the viewer's attention towards them. Canis~\cite{ge2020canis} is a high-level domain-specific language that allows users to select marks from charts and apply custom animations—e.g., a ``wheel'' effect on a donut chart—to mark units.

These prior works that tether data to either static or animated elements consider the graphic itself as a first-class object and what the user generates first in the workflow. In contrast, our work considers \textit{text} as the users’ first point of entry and recommends all subsequent bindings with respect to the context of this text. We then establish text-chart primitive-graphic bindings to support a greater breadth of data expression.


\subsection{Text-powered Data Stories}

When coupled with existing charts in a document, text can provide invaluable context for how a data story may be directed. Systems have thus been created to utilize this text to 1) highlight, 2) reformat, or 3) generate data visualizations. To guide users to corresponding charts when they are reading through a document, Elastic Documents~\cite{badam2019elastic} couples data tables with text that the user focuses on using a keyword-based matching algorithm to produce on-demand visualizations for whatever the user is reading. Automatic Annotation Synchronizing~\cite{lai2020automatic} additionally extracts visual elements from graphs using Mask R-CNN to automatically sync these visualizations to accompanying textual descriptions, which can be focused to highlight the graph. ChartText~\cite{pinheiro2022charttext} achieves the same automatic binding using a two-stage encoding method and can be used to add interactivity to documents. Kori~\cite{latis2022kori} both automatically suggests and allows users to manually generate references between text and existing graphs from a database as they type.

Alternatively, another strategy to redirect viewer attention more conspicuously is to rearrange the order of text and charts entirely. To improve the flow of data articles across the dynamic page layouts of different devices, VizFlow~\cite{sultanum2021leveraging} establishes text-chart links and reorganizes the text and charts based on each layout. ToonNote~\cite{kang2021toonnote} similarly reminds users of the bigger picture during data analysis by providing a toggle-able ``Comic View'' for computational notebooks.

In instances where visualizations may not exist to convey the exact message delivered by the text, systems may generate them instead, both with or without the presence of data. CrossData~\cite{chen2022crossdata} establishes text-data connections to help users retrieve, compute, and explore tables and charts during their document writing process. Similarly, DataParticles~\cite{cao2023dataparticles} links text and data, but specifically utilizes latent connections to help users iterate on animated unit visualizations that accompany the narrative.

Generative systems such as CrossData and DataParticles are most similar to our work. However, unlike current systems that focus on the retrieval of existing charts and then modify them to fit the user's data, we generate other data story-relevant assets such as static images and animations, and support interactions to combine them to better complement the message of the data story.

\subsection{Recommendations for Infographic Creation}

Since infographic authoring is often a multi-step and multi-platform process, recent recommendation systems have tried to automate or expedite various aspects of this process. For example, it is often challenging for novices to determine proper layouts and configurations for infographic assets, as different underlying semantic structures linking visual elements can lead to different stories conveyed to the user~\cite{lu2020exploring}. To ameliorate this challenge, Infographics Wizard~\cite{tyagi2022infographics} relies on a semi-automatic framework to recommend visual information flow layouts, visual groups, and connecting elements between assets. Similarly, Zheng et al.~\cite{zheng2019content} proposed a fully automated approach that uses input images and keyword-based summaries of input text to suggest magazine layouts. De-Stijl~\cite{shi2023destijl} \chirevision{and InfoColorizer~\cite{yuan2021infocolorizer}} recommend harmonic color palettes for novice users to assist them in quickly crafting design iterations.

Beyond layouts, other systems focus more on providing recommendations for the content of infographics directly. InfoNice~\cite{wang2018infonice} associates custom graphics with summarized data to transform unembellished charts into infographics with customized marks. \chirevision{ChartSpark~\cite{xiao2023let} embeds semantic context into existing charts using a text-to-image generative model.} InfoMotion~\cite{wang2021animated} converts static infographics into animated ones by producing a logical breakdown of components within the visualization. DataShot~\cite{wang2019datashot} automatically generates fact sheets and TypeDance~\cite{xiao2024typedance} generates typographic logos based on design priors from existing templates. Text-to-Viz~\cite{cui2020text} generates infographic content based on proportion-related statistics from statements by retrieving vector graphics from a database and arranging them according to a predefined list of 20 templates. Similarly, given a text-based prompt, Retrieve-Then-Adapt~\cite{qian2021retrieve} retrieves existing infographics from a database and transforms the content to match user-inputted data.

While our work also emphasizes infographic content generation, we do not rely on templates and source all assets from a centralized message. This workflow can thus provide greater creative autonomy for users while still ensuring that the main idea conveyed by the resulting design is maintained.

\begin{table*}[ht]
\small
    \centering
    \chirevision{
    \begin{tabular}{p{1.5cm} p{3.5cm} p{3cm} p{1.5cm} p{1.15cm} p{1.9cm}} 
        \toprule
        {Infographics Category} & {Definition} & {Authoring Method} & {Artifact Examples} & {System Examples} & {Components} \\
        \midrule
        \textbf{Statistical} & Relies on formal or stylized diagrams, charts, and graphs to summarize or highlight data & Create either standardized, annotated, or stylized visualizations & Posters, Maps & \cite{kim2017datadriven, xia2018dataink, yuan2021infocolorizer, ren2019charticulator, zhang2020dataquilt, wang2019datashot, cao2023dataparticles, pinheiro2022charttext, lai2020automatic, xiao2023let} & Text, Visualizations, Graphics\\
        \textbf{Mnemonic} & Relies on patterns, composition, and structure to highlight specific features and characteristics & Arrange text, symbols, or images in thematic layouts & Banners, Brochures & \cite{cui2020text, yuan2021infocolorizer, shi2023destijl, tyagi2022infographics, qian2021retrieve} & Text, Graphics, Symbols, Colors, Layout\\
        \textbf{Transitive} & Relies on either interactive or automatic transitions to convey sequences of events or operations & Design transitions within or between visualizations, images, and text & Slideshows & \cite{cao2023dataparticles, wang2021animated} & Text, Visualizations, Graphics\\
        \textbf{Directive} & Relies on spatial ordering to establish logical flow or direction & Sequentially arrange text, visualizations, and images & Data comics, Instructions, How-to's & \cite{kim2019datatoon, chen2022crossdata, kang2021toonnote, sultanum2021leveraging, latis2022kori, pinheiro2022charttext, lai2020automatic} & Text, Visualizations, Graphics, Symbols, Layout\\
        \bottomrule
    \end{tabular}}
    \caption{\chirevision{A summary of the main types of infographics, how and through which system they are typically authored, and a breakdown of their components.}}
    \label{tab:infographic_types}
    \Description{A table with six columns and four rows depicting the infographic category, the definition of that category, the authoring method for that category, example infographics, example authoring systems, and components that make up that category.}
\end{table*}
\normalsize

\section{Design Space}

\chirevision{To contextualize our proposed message-based infographic authoring paradigm, we first need to understand its existing design space. This process to distill design goals has been previously followed for other tools that focus on proposing new interactions to create visualization~\cite{chen2020augmenting,chen2019marvist,tong1912exploring}. Thus, we surveyed prior work on the authoring of visual data stories. To probe real-world usage scenarios of this design space, we also interviewed 2 design experts, one within academia and another in industry. Our derived insights are synthesized in the subsections below.}

\subsection{What are the different types of infographics and what do they contain?}
\label{sec:infographicTypes}

\chirevision{Based on prior work on infographics usage~\cite{siricharoen2013infographics, tarkhova2020infographics}, a breakdown of how different infographics are authored and what components they are composed of is located in Table \ref{tab:infographic_types}. Formally, an infographic is defined as ``a collection of graphic organizers that integrates different media in simple diagrams: text, images, symbols and schemas''~\cite{siricharoen2013infographics}. Images can be further broken down into visualizations that summarize data or data-agnostic graphics that convey information through visual metaphors. Similarly, a schema can be decomposed into the color palette that governs the media and the layout, or visual groups~\cite{lu2020exploring}, they are arranged in. Thus, an infographic can be considered a composition of \textbf{6 main primitive components: text, (data) visualizations, (non-data) graphics, symbols, color palettes, and layout}. We further dissect the primitive components to reveal the \textbf{functional components} that could be provided by an authoring system. First, graphics and symbols can be derived from both (C1) \textbf{raster and vector graphical formats}. Then, visualizations as data charts and annotations could be derived from (C2) \textbf{declarative grammars} and (C3) \textbf{queries on the chart data}. Layouts are attributed to different (C4) \textbf{templates of visual groups}~\cite{lu2020exploring}. Finally, (C5) \textbf{color palettes} and (C6) \textbf{text} can be self-contained collections of hues or fonts, respectively. An authoring system could provide these items as primitives for further combinations and editing to construct an infographic.}

\subsection{How can a \textit{tangible} key message support the authoring of infographic components?}

\chirevision{Given the breadth of infographic components, supporting users to create them all from scratch is not an easy task, especially when an effective infographic requires the author to be clear and succinct in what they wish to convey~\cite{murray2017maximising}, while juggling multiple visual elements on the page. We identify an opportunity here where many components originate from the same concept: a ``key message'' the designers want to convey. While there are scenarios where analysts performed an initial exploration of data first, they commonly use such key messages to pass on what they want to designers.}
\chirevision{Thus, we propose a workflow that generates these components by employing an explicit singular text-based key message---an ``epigraph'' that alludes to the contents to come. Such an epigraph would contain rich semantic information that can be broken down into themes, sub-phrases, and individual words that could be treated separately as prompts for component primitives. Themes could suggest potential \textit{color palettes} by extracting discrete words and combining them harmoniously. Sub-phrases can provide abstractions for \textit{visualizations}. Individual words can be used to retrieve \textit{graphics} and \textit{symbols}. The key message itself or its constituents can be added as \textit{text}. Note that while \textit{layout} is a primitive component that has been generated in prior systems~\cite{tyagi2022infographics, qian2021retrieve, cui2020text}, we deliberately exclude layout and font recommendations as they usually require extrinsic input such as user style or templates that could not be systematically extrapolated from a message. Thus, we expect that a self-contained key message would provide the following components: \textit{graphics (C1), visualizations (C2, C3), color palettes (C5), and the text content (C6)}.
}

\begin{figure*}[t]
  \includegraphics[width=\textwidth]{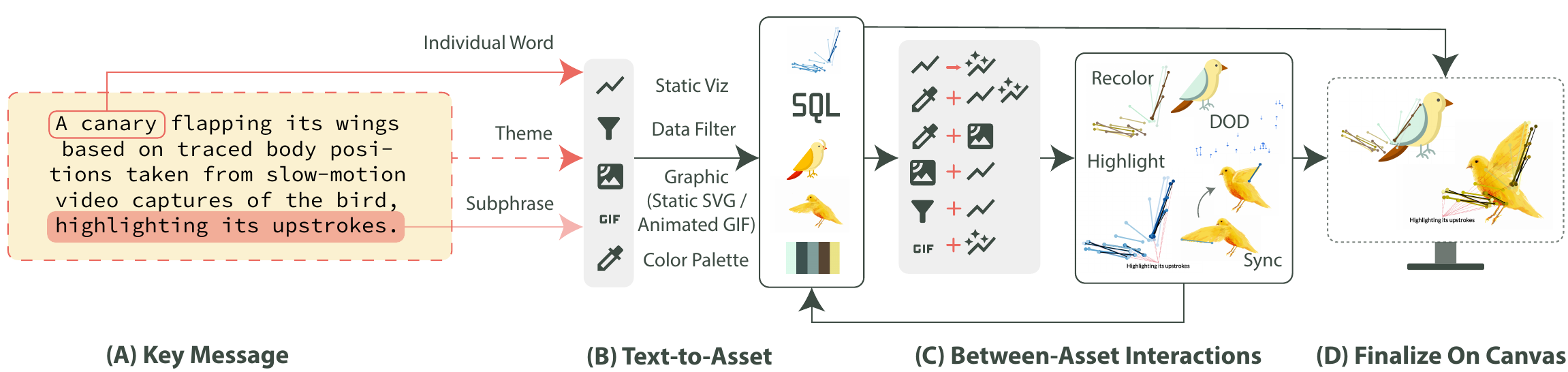}
  \caption{The complete pipeline from a text-based key message to infographic elements. It involves selecting text chunks from a key message (A), using these chunks to recommend different types of assets (B) such as visualizations, data filters, graphics, and color palettes, merging different combinations of the generated assets (C), and fine-tuning the configurations on a canvas (D).}
  \label{fig:authoring-workflow}
  \Description{A flowchart showing how a key message is converted to an infographic. The first node in the flowchart showcases a key message with different types of text (individual words, sub phrases, or the entire message) highlighted. Arrows from these text chunks feed into the next node, a list of icons showing different asset types. An arrow is drawn from them to a list of assets generated. From this list, two arrows are drawn–one straight to the final infographic and another to a list showing icons of the asset types added to each other, representing how assets can be combined. From this list, another arrow is drawn to the outcomes of these asset combinations. Finally, two arrows are drawn from these combinations–one back to the list of assets generated and another to the final infographic.}
\end{figure*}

\subsection{How can components be further composed to form varied types of infographics?}
\chirevision{We additionally want to support ways to craft each \textit{type} (Table \ref{tab:infographic_types}) of infographic, which is determined by the purpose assigned to each component and how they interact with each other. For example, a statistical infographic may necessitate annotated or stylized visualizations; the former requires integrating text into the visualization as annotations, while the latter embeds graphics as glyphs within the visualization. Conversely, mneumonic infographics are not data-based, but rather rely on the layout of text, graphics, symbols, and color to convey their message. Transitive infographics add animations to visualizations or graphics to illustrate sequential events, while directive infographics convey the change through spatial ordering, organized either instructional symbols via the layout of media. In authoring these different types of infographics, the same component can serve multiple functionalities--text can be treated as an annotation to or a title summarizing the visualization. The message conveyed can also be represented as different components--it could be explicitly displayed as a titular banner or left latent as an implication from a graphic. Thus, allowing the recommended components to be combined in a unified UI allows the re-purposing of components into the diverse roles necessary for each infographic type. This UI should also allow users to manually control layout, which cannot be automated solely from a key message.}

\subsection{Design Goals}

\chirevision{Our tool is motivated by the existing design space of infographics authoring and introduces a top-down approach based on the assumption that the author has a key message in mind. Specifically, it intends to achieve the following:}

\chirevision{\begin{itemize}[leftmargin=3mm]
    \item \textbf{G1}: Support natural ways to interact with and extract all relevant information from text towards component creation (message to component recommendations). 
    \item \textbf{G2}: Accommodate higher-level abstractions to compose components so that different categories of infographics are supported (component re-purposing and combination).
    \item \textbf{G3}: All interactions should occur in and can be fine-tuned within a unified system to reduce context-switching (human-in-the-loop to handle components and modalities that cannot be achieved via automation in G1 + G2).
\end{itemize}}

\section{Epigraphics System}

\chirevision{Our proposed tool, \systemname, is situated in the larger space of storytelling tools~\cite{li2023where, lee2015more}, specifically within the planning and implementation stages of the workflow. It also sits at a niche between systems that provide complete low-level fine-grain control desired by domain experts~\cite{xia2018dataink, cao2023dataparticles} and generative systems that completely automate the text-to-design process~\cite{cui2020text} with its AI creator and human optimizer model~\cite{li2023where, li2023ai}. It aims to empower infographic authoring through a key message-sourced approach to compose and combine component primitives. For clarity, we will use the term \textit{component} to refer to the primitives that make up a general infographic and the term \textit{asset} to refer to the primitives that can be extracted from the key message. The two sets mostly overlap, with some differences we will discuss in Section \ref{sec:assetRec}.}

\subsection{Pipeline Overview}

\noindent In our system, text is the primary element users interact with to generate different assets for infographic design. The pipeline that describes this workflow is depicted in Figure \ref{fig:authoring-workflow}, which uses a toy example of a user-inputted sentence, one about a canary's wingspan, to illustrate how assets may be recommended. Assuming that the relevant dataset is already imported, the user begins by \chirevision{indicating} what components of the key message they wish to use for their source input \chirevision{by brushing over the text with their cursor}. They then specify the asset type they want to see, which the system uses to generate a ranking of the selected asset type most relevant to the source input (\textbf{G1}). From the list, the user can then click on each asset to directly add it onto the canvas, or further modify it even more by combining the generated assets (for example, using the color palette to recolor the SVG results of the canary) (\textbf{G2}). Finally, they can rearrange all the assets on the same canvas, making more modifications if necessary, to produce a finalized infographic (\textbf{G3}).

\subsection{Data Preparation and Generative Models}

Recently, large-language models (LLMs) have attained high levels of generality over a wide range of tasks due to their scale and attention-based architectures~\cite{kaplan2020scaling}. This makes them ideal candidates for text-to-asset and text-to-design generation, \chirevision{especially if the LLM is used to produce multiple iterable segments within a master draft~\cite{sultanum2023datatales}}. However, since these models do not actually understand the prompts they are provided with like a human would~\cite{webson2021prompt}, prompts can be engineered to better assist the model in converting natural language instructions into desired outputs~\cite{liu2022design, white2023prompt}. From CSV files containing the dataset, we extract meta-information such as column names, a high-level summary of the data, and unique categorical values to provide additional context to LLMs. For the graphics, we extract captions for each image using a Visual Question Answering (VQA) model~\cite{antol2015vqa} by asking \textit{``what does the image show?''} We then use a family of generative models suitable for each type of resource. Specifically, \verb|GPT-3.5|~\cite{gpt3-5} is utilized for text completion, Sentence-BERT~\cite{reimers2019sentence-bert} for embedding extraction, BLIP~\cite{li2022blip} for image caption generation, and Adobe Firefly\footnote{\url{https://www.adobe.com/sensei/generative-ai/firefly.html}} for text-to-image generation to extract color schemes. 

\subsection{Recommending Infographic Components (Text-to-Asset)} \label{sec:assetRec}

\chirevision{Recall from Section \ref{sec:infographicTypes} that a generic infographic may contain 6 primitive components: text, data visualizations, graphics, symbols, color palettes, and layout. Our system uses one of these components, text, to recommend all other components except for layout. We note that symbols are a subset of graphics~\cite{siricharoen2013infographics} that can be generated based on specific text. We further introduce an additional asset, data filters, that can be extracted from the text and used to refine and re-purpose other components such as visualizations and text. Thus, the full set of primitive assets \systemname supports includes (static) visualizations, data filters, static and animated graphics (including symbols), and color palettes.}

\subsubsection[]{\includegraphics[width=0.02\textwidth]{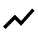}\textbf{Static Visualizations}}

\chirevision{Text could be a strong signal for what and how a summary of the data could be displayed to convey the desired message. Assuming an appropriate dataset has been imported and the user has indicated their text of choice}, we use \verb|GPT-3.5| to extract no more than 5 most relevant \textit{column names} to the user-provided text. These columns are then converted into intent grammar using Lux~\cite{lee2021lux}. 
Lux's intent language allows partial specification based on CompassQL~\cite{wongsuphasawat2016towards}. 
\chirevision{
Specifically, it only requires specifications for data aspects of interest (i.e. column names or data filters) and does not need inputs for visualization encodings.
This reduction of columns, a common task in data exploration~\cite{fernandez2018aurum}, contracts the search space of visualization specifications so that the recommendation is less likely to depend on the chart recommendation engine.
}
The 5 (or fewer) columns are further broken down into subsets of 2 columns (scatter plots, line charts, bar charts), 3 columns (charts with an additional colored legend), or aggregated/binned (histogram, heatmap) to generate the most common chart types. These output charts are ranked based on the number of relevant columns involved, then converted to Vega-Lite specifications \cite{satyanarayan2017vegalite}, rendered as SVGs, for the user to choose from.



\subsubsection[]{\includegraphics[width=0.02\textwidth]{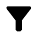}\textbf{Data Filters}}

\chirevision{Text could be used to indicate that the user only wants to operate on a subset of the data}. In these cases, the prompts are fed into \verb|GPT-3.5| to output SQL queries for the dataset. For example, a dummy sentence such as ``The Lakers vs Detroit finals in 2004 was particularly exciting'' will be converted to 
\begin{verbatim}
    SELECT * FROM df 
    WHERE team_name = 'Los Angeles Lakers' 
    AND opponent = 'DET' AND season = '2003-04' 
    AND period = 2 AND playoffs = 1 
    ORDER BY date LIMIT 10
\end{verbatim}
The query is applied to the dataset and the filtered data is returned as a table for the user to interact with. The table can either be used independently to generate new visualizations, be used to highlight existing ones as an overlay, or used to generate annotations for specific data points, which we will discuss in Section \ref{sec:betweenAssets}.


\subsubsection[]{\includegraphics[width=0.02\textwidth]{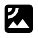}\includegraphics[width=0.02\textwidth]{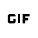}\textbf{Static \& Animated Graphics}}

\chirevision{While designers traditionally look up relevant images to import into an infographic design, sourcing these assets directly from the text can streamline the process by keeping it centralized within the authoring tool.} Our system relies on SVGRepo\footnote{\url{https://www.svgrepo.com/}}, an open-licensed database for SVGs, as the source for its static graphics. For each image, we generate captions and extract sentence embeddings from them. Similarly, we also obtain embeddings from the user-provided text. After computing cosine similarities between the embeddings for the captions and user input, we rank the scores and return the top 20 SVGs as recommendations for the user. Recommendations for animated graphics are returned as GIFs. GIFs are generated similarly by average pairwise cosine similarities between the user input and each frame in the animation. 


\subsubsection[]{\includegraphics[width=0.02\textwidth]{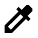}\textbf{Color Palettes}} \label{sec:colorGen}
\chirevision{Finally, while the text structure itself cannot denote a comprehensive color palette, specific keywords can suggest potential colors that make up one. The users' text of choice is broken down into such keyword fragments, each of which is fed into} the text-to-image module of Adobe Firefly to produce multiple relevant images. \chirevision{From each image}, we extract the color profiles by computing 5-bin color histograms from each image. The histograms are then compiled into color palettes with color sorted by luminosity.

\begin{figure}[t]
  \includegraphics[width=1.0\linewidth]{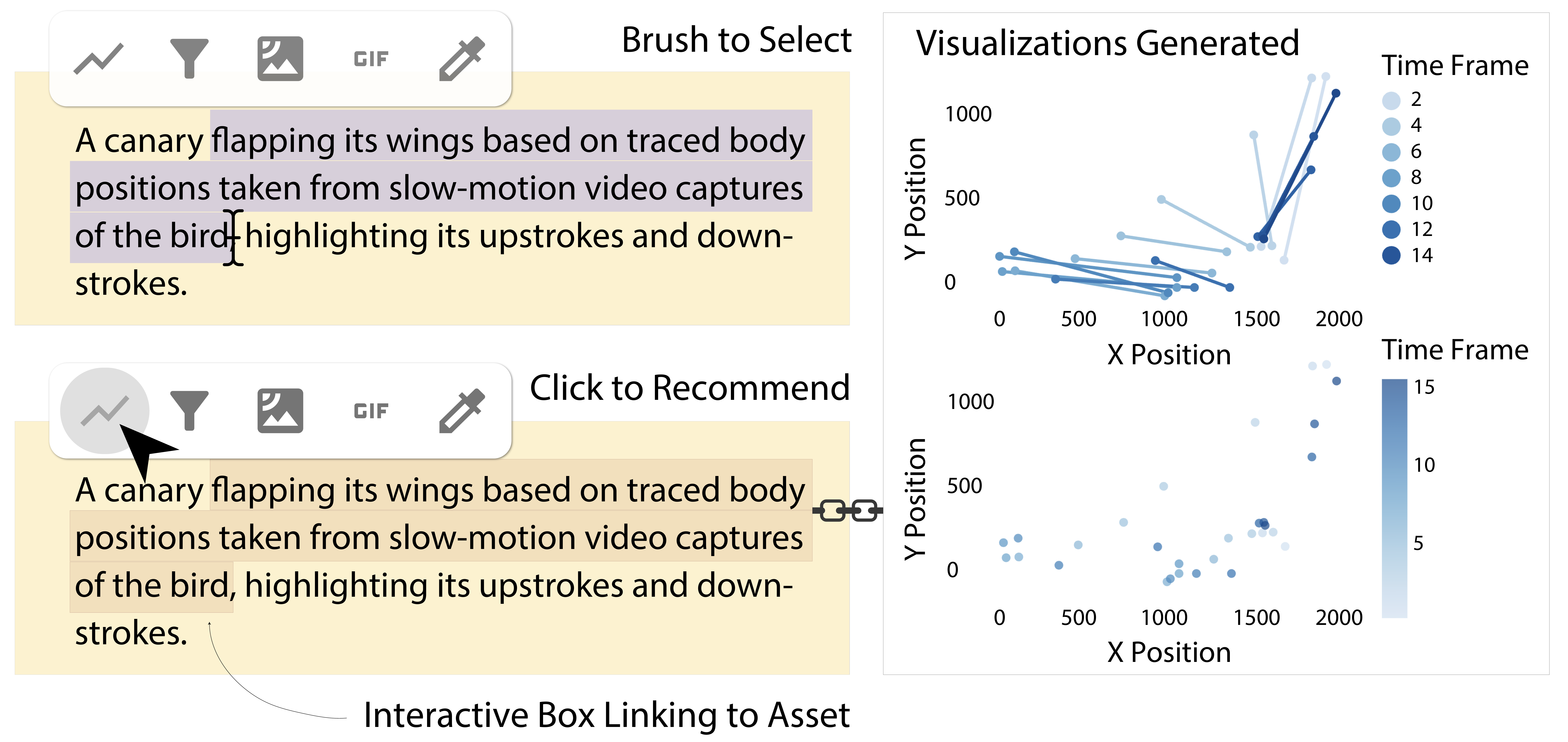}
  \caption{When the user brushes over a chunk of text, a pop-up with icons representing potential types of asset recommendations appears. After the user clicks on an icon, the asset is generated and automatically linked to the text chunk.}
  \label{fig:text_editor_brush_select}
  \Description{Two text boxes containing the same key message next to a box contain three visualizations. In the first text box, a cursor brushes over a block of text and a popup with 5 icons indicating visualization, data filter, static graphic, animated graphic, and color palette shows up. In the second text box, the cursor clicks on the visualization icon and an interactive box is formed around the block of text. The box with visualizations contains a connected scatterplot and a scatterplot, both linked to the interactive text box via a chain icon.}
\end{figure}

\begin{table*}[t]
\small
    \centering
    \begin{tabular}{lllll}
        \toprule
        {Configuration} & {Supported Asset Types} \\
        \midrule
        Hide/show axes \& legends  & static visualizations, animated visualizations, data-oriented drawings\\
        Add animation & static visualizations\\
        Manual recolor & visualizations, graphics, text\\
        Change opacity & visualizations, graphics, data-oriented drawings, highlights, text\\
        Change thickness/size & visualizations, data-oriented drawings, annotation lines of highlights, text\\
        Change style/pattern & annotation lines of highlights, text\\
        Change frame delay & animated visualizations, animated graphics\\
        \bottomrule
    \end{tabular}
    \caption{A summary of all possible configurations for each asset type.}
    \label{tab:layer_configs}
    \Description{A table with two columns showing the asset configurations possible and what asset types support them.}
\end{table*}
\normalsize 

\subsection[]{Component Re-purposing and Combination (Between-Asset Interactions)}~\label{sec:betweenAssets}
Once the text-based assets are generated and added to the canvas, the system provides further options to combine or refine them \chirevision{to accommodate the intentions each component serves in different types of infographics (Table \ref{tab:infographic_types}). While this is not a comprehensive list of \textit{all} possible asset-asset interactions, 
most extended combinations could be achieved based on the following pairwise combinations from the core assets.} Note that these interactions are commutative operations where the order of the assets does not matter, 
and the outputs of each combination can additionally be combined with another (e.g. a highlighted visualization can be recolored based on a color palette or synced with an animated graphic). 

\subsubsection[]{\includegraphics[width=0.02\textwidth]{assets/show_chart.pdf}Static Visualizations $\rightarrow$ \textbf{Animated Visualizations}}

Certain datasets may also contain temporal attributes that can be animated \chirevision{to demonstrate a change over time}. Based on a visualization of interest with an associated dataset, we prompt \verb|GPT-3.5| with \textit{``output the columns with time-oriented words''} to extract these columns, which are then presented in a drop-down menu for the user. Once the user selects a column, we convert the unique values of that column into a set of ordered keys that define each frame in the animation. The dataset of the visualization is then filtered for each key, resulting in a GIF that loops indefinitely over each unique time-oriented column value. The user can also freely control the frame delay and thus the speed of the animation.

\subsubsection[]{\includegraphics[width=0.02\textwidth]{assets/colorize.pdf} Color Palettes + \includegraphics[width=0.02\textwidth]{assets/show_chart.pdf} Visualizations, \includegraphics[width=0.02\textwidth]{assets/satellite.pdf}\includegraphics[width=0.02\textwidth]{assets/gif.pdf} Graphics $\rightarrow$ \textbf{Recolor}}

Once a desired color palette is selected from the list of recommendations, the user can click on any SVG, GIF, or visualization to map the palette onto the colors of that asset. The color palette from the visualization or graphics is extracted via color histograms in a similar way as Section~\ref{sec:colorGen}. Given two histograms, the system then transfers the colors by mapping them to minimize the Earth Mover Distance. Note that regardless if a visualization has a categorical, diverging, or linear color scheme, the resulting colors will preserve these properties.

\subsubsection[]{\includegraphics[width=0.02\textwidth]{assets/satellite.pdf} Graphics + \includegraphics[width=0.02\textwidth]{assets/show_chart.pdf} Visualizations $\rightarrow$ \textbf{Data-oriented Drawings}}
We define a data-oriented drawing (DOD) as a stylized visualization that incorporates custom imagery as glyphs, similar to the outputs of Charticulator \cite{ren2019charticulator} or DataQuilt \cite{zhang2020dataquilt}. To create a DOD, the user can select any existing visualization on the canvas with a categorical colored legend. Then, after selecting what images they want to replace each legend value with from the list of recommended graphics, the visualization is automatically replaced with glyphs. This combination works on scatterplots, bar charts, and line charts with markers. 

\subsubsection[]{\includegraphics[width=0.02\textwidth]{assets/filter_alt.pdf} Data Filters + \includegraphics[width=0.02\textwidth]{assets/show_chart.pdf} Visualizations $\rightarrow$ \textbf{Highlighted \chirevision{Visualizations + Annotations}}}
Given a data filter, there are two ways to modify a visualization. If the visualization is a result of aggregated data, the result is the same visualization but abstracted from less data. If the data has not been aggregated, the result is a selection of the current encoding presented as an overlay. We additionally provide annotation-like lines \chirevision{containing the initial text chunk that prompted the highlight} which point to the filtered data in the visualization.


\subsubsection[]{\includegraphics[width=0.02\textwidth]{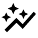} Animated Visualizations + \includegraphics[width=0.02\textwidth]{assets/gif.pdf} Animated Graphics $\rightarrow$ \textbf{Sync}}

When animated visualizations are added to the canvas in conjunction with animated graphics, the user may wish to sync the animations to create greater unity within the infographic, provided that both have the same number of frames. If not, we either trim the animated graphic or the animated visualization to ensure their frame count is a multiple of the other. Then, we sync their timings by mapping the frame delays of the animated visualization to the graphic and resetting both animations to start simultaneously. 

\begin{figure*}
  \includegraphics[width=\textwidth]{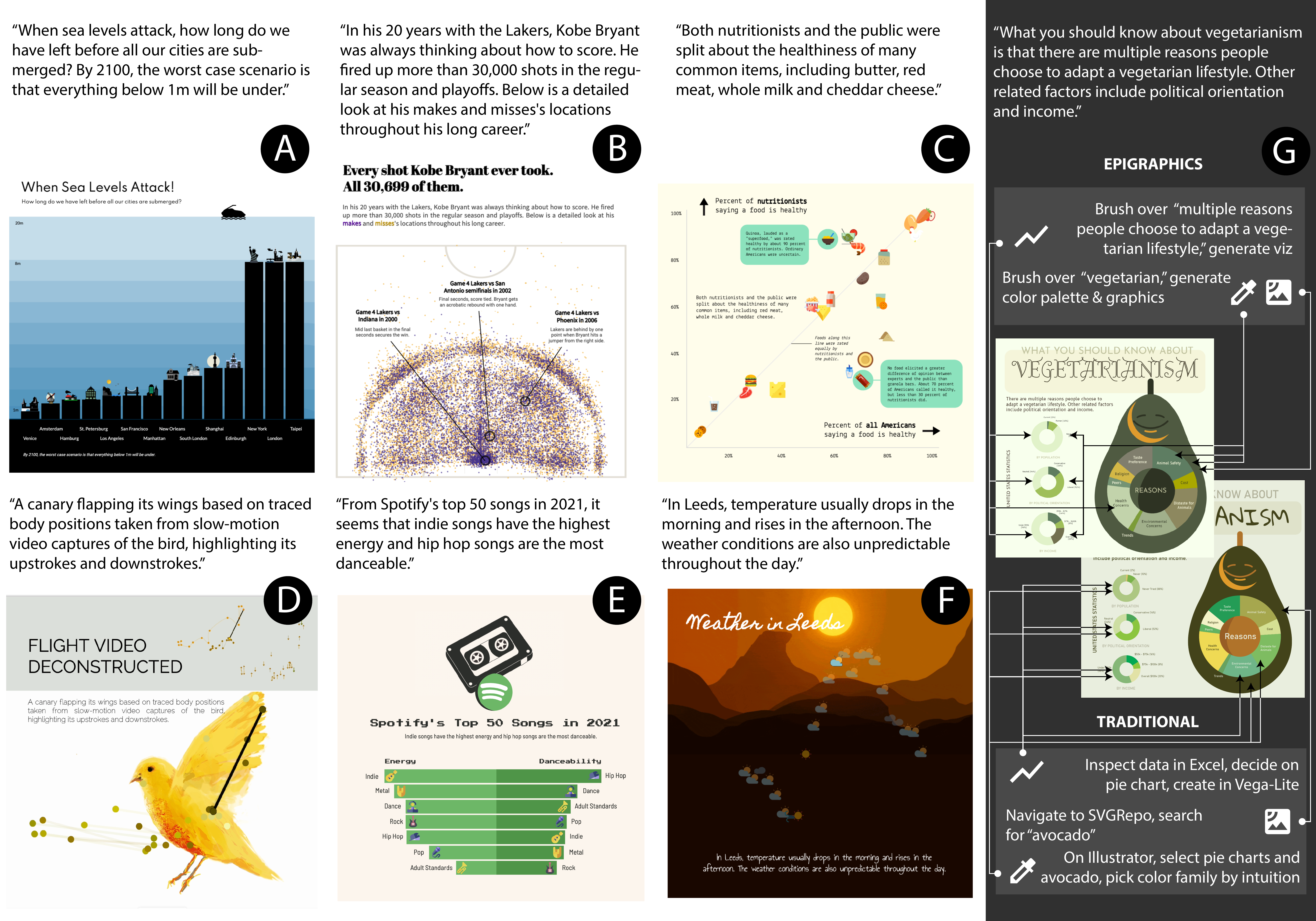}
  \caption{A gallery of infographics created using \systemname with the corresponding epigraphs used to generate them. \chirevision{A~\cite{mccrorie2016infographics}, B~\cite{fox2016every}, C~\cite{quealy2016sushi}, and D~\cite{lutz2014flight}} showcase recreations inspired by existing infographics, while E and F are originals based on open-source datasets. \chirevision{G~\cite{zuniga2014what} is also a recreation comparing what can be produced using our system (top) versus a traditional approach (bottom) with labels explaining their workflows for each asset type.}}
  \label{fig:exemplar_gallery}
  \Description{A gallery of infographics arranged in 2 rows of 4 with the epigraphs used to create them on top. In order: A) A stylized bar graph with cities on top of each bar and a gradient of blue in the background from light (top) to dark (bottom). B) A scatterplot overlaid on top of a basketball court, with three points annotated. C) Icons of food items representing each data point of a scatterplot. Icons representing quinoa and granola bars are highlighted. D) A connected scatterplot overlaid on top of a graphic of a canary. E) A mirrored bar chart with icons representing each genre at the tips of the bars. Graphics of a record player and Spotify icon are above this chart. F) Icons representing weather conditions are highlighted and match the position of the sun in the background. G) Two similar infographics of four pie charts, one of which is contained within an avocado.}
\end{figure*}

\subsection{Authoring Interface}

The authoring interface, depicted in Figure \ref{fig:user-interface}, is a web application built using Node.js and Next.js. API calls for text-sourced recommendations are sent to a Flask backend. Before interacting with the interface, the user can either select a preset dataset provided using the dropdown or upload their own CSV to explore that data.

\subsubsection{Text Input Panel}

The text input panel (Figure \ref{fig:user-interface}A) is a rich editor where the user can enter the key message of the infographic or any other text that they wish to use to recommend assets with. 
Initiating an infographic asset recommendation involves using the cursor to brush over a phrase of interest to select it (Figure~\ref{fig:text_editor_brush_select}). This brings up a panel containing icons for the potential assets available to be recommended (\includegraphics[width=0.02\textwidth]{assets/show_chart.pdf}\includegraphics[width=0.02\textwidth]{assets/filter_alt.pdf}\includegraphics[width=0.02\textwidth]{assets/satellite.pdf}\includegraphics[width=0.02\textwidth]{assets/gif.pdf}\includegraphics[width=0.02\textwidth]{assets/colorize.pdf} from Section~\ref{sec:assetRec}).
 Clicking on the icons sends an API call to fetch the corresponding asset, which is then appended to the bottom of the recommendations list. The selected text chunk is automatically linked to the generated list of assets, represented as an interactive box around that chunk in the text editor. Users can then click on the box to filter and find the corresponding assets easily when the recommendation list grows longer.

\subsubsection{Recommendations List}

The recommendations list (Figure \ref{fig:user-interface}B) contains a scrollable history of asset recommendations requested by the user, labelled with the text they are sourced from. For visualization and graphic recommendations, clicking on any element will generate an SVG (or GIF if animated) instance of that asset on the canvas. Data filters are represented as selectable tables and toggling a table row while a visualization on the canvas is selected will create annotated highlights over that visualization. Clicking on a color palette with a visualization or graphic on the canvas selected will recolor that asset based on the palette color schemes. There is also a list of tabs above the recommendations list, which the users can use to filter the list based on their desired asset category.

\subsubsection{Canvas}

The canvas (Figure \ref{fig:user-interface}C) is a space for users to freely manipulate the assets \chirevision{, make final adjustments, and author aesthetic layouts if desired}. It supports basic editing functionalities like the ability to add text, change text font, change color, undo/redo, change asset opacity, lock assets, move the depth of assets forward/backward, and download. Users can also move, rotate, and rescale added assets using direct manipulation. Since animations are started once they are added to the canvas, we also include an additional function to reset all the animation timings so that they can start at the same time if desired. 

\subsubsection{Layers System + Configurations}

Additional fine-tuning of the added assets is possible in the layers system (Figure \ref{fig:user-interface}D). A summary of all possible configurations can be found in Table \ref{tab:layer_configs}. For example, users can toggle the visibility of visualization properties such as axes and legends, as well as select the time-oriented column to animate over for animated visualizations. All assets can be recolored by manually mapping one color in the asset to another. Annotation lines can be adjusted for thickness, color, scale, and start/end head patterns. Text anchor positions to the annotation lines can be adjusted. Frame rates for animations can also be adjusted.
\section{Gallery and Case Studies}\label{sec:cases}

\systemname aims to facilitate the rapid, focused creation of message-based infographics. \chirevision{Thus, it should support users in both creating an infographic from scratch and recreating the ideas of existing ones given a meaningful message}. We created a gallery to demonstrate the expressiveness of the system in accommodating these goals (Figure~\ref{fig:exemplar_gallery}). \chirevision{Five of these examples are inspired replications (A~\cite{mccrorie2016infographics}, B~\cite{fox2016every}, C~\cite{quealy2016sushi}, D~\cite{lutz2014flight}, G~\cite{zuniga2014what})} that come from news articles, posters, online blogs, and other story telling mediums that contain rich graphics, visualizations, and textual descriptions. Two of these (E, F) are original infographics where we came up with our own messages after exploring public datasets.

\chirevision{To illustrate the workflow and highlight the capabilities, strengths, and limitations of our system, we also present two case studies. Case Study 1 is a walk-through for producing infographic D, showing the mechanism that a bird uses to fly, while Case Study 2 directly compares workflows with and without \systemname for infographic G. The workflow without \systemname was completed using Vega-Lite to manually compose the visualization, SVGRepo to source graphics, and Illustrator to combine them. This ensures that the assets in both workflows are the same to minimize confounding factors in the comparison.}


\subsection{Deconstructing the Flight of a Canary (D)}
\label{sec:case_study_one}
\chirevision{After importing the dataset into the system, we entered the key message (Figure \ref{fig:exemplar_gallery}D Top) into the text panel. Inspired by the original infographic~\cite{lutz2014flight}, we noted that we needed an animated graphic of a canary and an visualization of the wing positions animated over each time frame overlaid on top, and looked for opportunities to extract this information from the text (G1). For the graphic, we brushed over the text ``canary flapping its wings'' to generate potential GIF candidates for the animated canary, and selected a yellow bird flapping its wings to add to the canvas. Next, we brushed over the text ``wings based on traced body positions'' to view potential visualization candidates. The returned 20 options included a mixture scatter plots, bar charts, and line charts for different data column values such as x position, y position, time frame, wing type, and wing stroke direction. We tried to narrow down these options by adding ``as an animated line graph,'' but this did not work. Instead, we found that adding \textit{semantically meaningful keywords} as such ``over time'' was more effective in trimming the recommendations by reducing the number of columns returned. That is, the tool made it particularly easy to generate a breadth of exploratory assets that can be filtered through a more refined qualitative message, but proved more difficult in executing explicit, technical commands that an AI chatbot would normally take.}

\chirevision{After scrolling through the options, a connected scatter plot that plotted x against y position with time frame as the legend was deemed the most appropriate. To really illustrate how the wingspans evolved as the bird flapped its wings (G2), we added animation in the configurations panel, which brings up a drop-down of time-related columns in the visualization. In this case, there was only one option, the time frame. By selecting this, the static visualization was automatically converted to an animation overlay where the connected line of the scatter plot cycles through the x and y positions at each time frame. Next, we wanted to modify the scatter point colors to better match that of the bird. To do this, we brushed over ``canary,'' which returned an array of palettes, one of which was a collection of yellows that we applied to the visualization. We then synced the animations of both the canary and the visualization. For final touches (G3), we added the original key message as a header, a title, and modified their fonts to complete the infographic. Here, we note that although the visualizations and graphics could be modified for color and opacity, it was not possible to change their overall style (i.e. make it painterly, like a charcoal sketch, or holographic-like as in the original infographic~\cite{lutz2014flight}). Future work that expands upon canvas capabilities or incorporates \systemname as a plugin into existing graphical editors can expand the diversity of its visual outcomes.}

\subsection{Unraveling Statistics for Vegetarianism (G)}
\chirevision{In both workflows, the key message to be conveyed about what one should know about vegetarianism was made prominent--in \systemname, it was always displayed in the text panel while in the traditional workflow, it was explicitly written as a large banner on the working canvas of Illustrator. To generate desired visualizations, \systemname adapts an exploratory approach similar to Case Study 1, where relevant pie charts were picked from a collection of other chart types that combined the most relevant dataset columns after brushing over text chunks. For graphics, specifically the image of the avocado which was inspired by the original infographic~\cite{zuniga2014what}, \systemname generated options for vegetables after we brushed over ``vegetarianism.'' However, an avocado was not a vegetable, so we had to modify the text panel and type ``avocado'' directly. To color the pie charts and avocado, we brushed over ``vegetarianism,'' and applied an earthy green color palette to all the assets.}

\chirevision{Conversely, for the traditional approach, we first inspected the dataset to see how it could be plotted. Then, we decided to create pie charts and wrote scripts to sum the data rows and create Vega-Lite grammars. The visualization was exported as an SVG and imported into Illustrator. For the graphic, we directly looked up ``avocado'' on SVGRepo, and selected an avocado image we desired, downloaded it, and added it to Illustrator. However, recoloring each element on the canvas was more tedious. First, we thought about what a ``vegetarian color palette'' meant. Then, we selected each asset, which had its own color scheme, and manually remapped it using the `recolor' functionality on Illustrator. Some assets were recolored more than once to balance the overall visual coherence.}

\chirevision{Given the replication task, the components of \systemname did not necessarily influence \textit{what} was produced in the end, but rather improved the efficiency of the workflow by reducing context-switching as all operations could be performed in one system. \systemname also \textit{helped reason} about what potential chart types, images, and color palettes were possible by providing options sourced from the message you intend to convey, whereas that decision-making process was completely left to the user alone in the traditional workflow. However, this can backfire as system reason leads to undesired results, such as the ``vegetarianism'' $\rightarrow$ ``avocado'' instance, but re-writing the text can quickly resolve this issue.}
\section{Usability Study}

\chirevision{To investigate how \systemname's text-first approach may direct, or redirect, designers' mental models, we also analyze the 1) intermediate and final visual artifacts created by the user and 2) their interaction patterns and verbalized thought processes during the authoring process.}
We first ran pilot studies with two participants. They were given a dataset and asked to create infographics with the system with their own text-based messages or sketches. During this process, they expressed that it was difficult to come up with either one after just looking at a table because it was hard to gain insights from the data or come up with a story to tell. In fact, they treated the system as a data analytics system and spent most time exploring the data using the recommended visualizations.
To avoid the deviation of purposes, we provide fixed messages for all participants in the final user study so that they could focus on the authoring experience instead of learning the datasets.

\subsection{Participants}

Ten users (6 female, 4 male), recruited via snowball sampling, participated in the usability study. They range from 19 to 56 years old ($\mu=28.6, \sigma=10.3$) and have varying levels of design (2 beginner, 2 novice, 3 intermediate, 2 advanced, 1 expert) and programming (1 novice, 2 intermediate, 5 advanced, 2 expert) expertise. Half of them have never read or written articles that discuss data, while the other half interacted with data articles regularly. Most of them, except for one user, have created visualizations before. However, only two participants frequently create infographics, while the rest reported to design them rarely. When they do, however, they cited using software such as Microsoft Excel, Figma, Adobe Illustrator, Adobe Express, and Canva to create them. 

\subsection{Study Protocol}

Participants were randomly divided into two groups that were given two different public domain datasets pulled from Kaggle. The first dataset contains weather conditions over a day in Leeds\footnote{\url{https://www.kaggle.com/datasets/muthuj7/weather-dataset}}, England, while the second dataset contains the 2021 top 50 Spotify songs and their acoustic properties\footnote{\url{https://www.kaggle.com/datasets/equinxx/spotify-top-50-songs-in-2021}}. All other conditions between the groups were kept identical. The study lasted a total of 60 minutes, and can be broken down into the following components:

\subsubsection{Introduction (10 minutes)}

Participants filled out a preliminary questionnaire about their prior expertise in design and programming, how often they read/write articles with data, create visualizations, and create infographics. They were then instructed to pretend to be a visualization graphic designer who was tasked with creating an infographic based on a single text-based message from a client. For inspiration, the participant was also shown three example infographics in case they did not have much prior experience viewing or creating them. 

\subsubsection{Sketching Task (10 minutes)}

Based on the text-based message, the participants were asked to sketch out their ideas for their infographic on a digital canvas. The purpose of this task was to understand their immediate first impressions of what they wanted the infographic to look like after reading the message. In this process, they were reminded that an infographic can include text, images, animations, and visualizations, but they are free to use (or not use) these elements in any combination they so desired \chirevision{and} encouraged to refer back to the message at any point during this process. 

\subsubsection{Infographic Creation Task (25 minutes)}

Participants were walked through the \systemname system via two demo videos that showcased how to create an infographic from text with two different \chirevision{datasets}. They also demonstrated the capabilities of the system, including the types of assets and the between-asset interactions that could be generated. The participants were encouraged to ask any questions they may have at this time. After the walk-through, participants were instructed to navigate to the \systemname interface and create their own infographic using the tool. The facilitator reminded them that they could either copy and paste the message the client gave them earlier directly or enter their own into the text panel. They were also encouraged to modify the message freely or use their initial sketch (or not) to achieve the vision they wanted. During this process, they were instructed to think aloud and ask if they had any questions. This task was considered finished whenever the participant felt that the infographic fulfilled the message or until 25 minutes were up. 

\begin{figure}{t}
  \includegraphics[width=\linewidth]{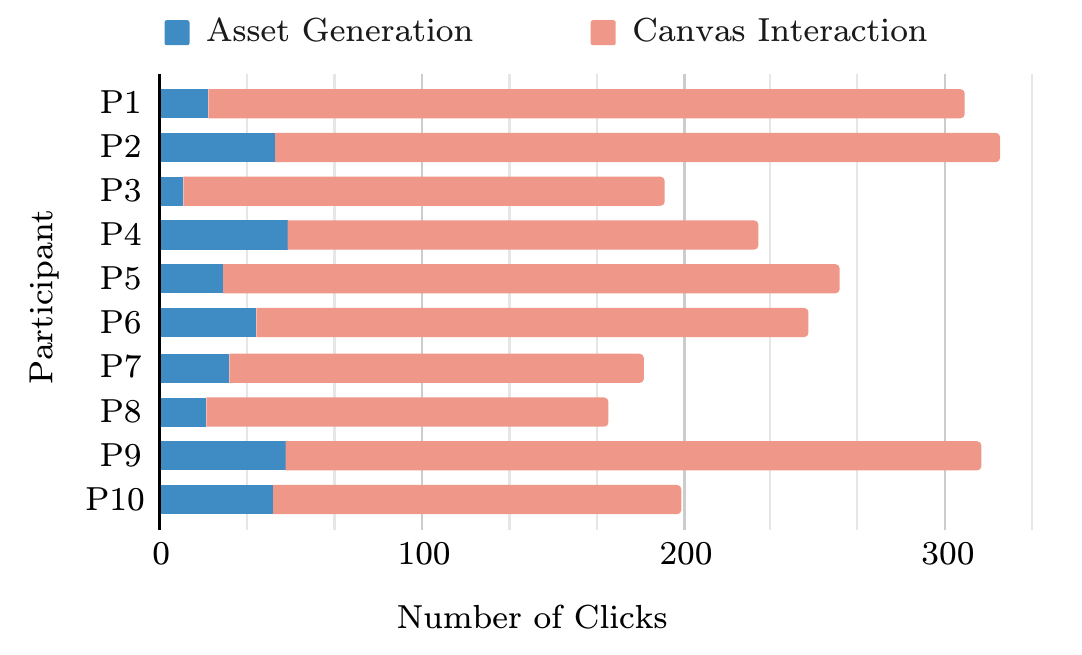}
  \caption{The distribution of number of mouse clicks spent on asset generation and interacting with the canvas for each participant. All participants spelled the bulk of their clicks on manipulating assets on the canvas.}
  \label{fig:clicks_per_participant}
  \Description{A stacked bar chart showing the proportion of asset generation vs canvas interaction clicks for each participant. Asset generation is significantly less compared to canvas interaction for all participants.}
\end{figure}

\begin{figure}[t]
  \includegraphics[width=\linewidth]{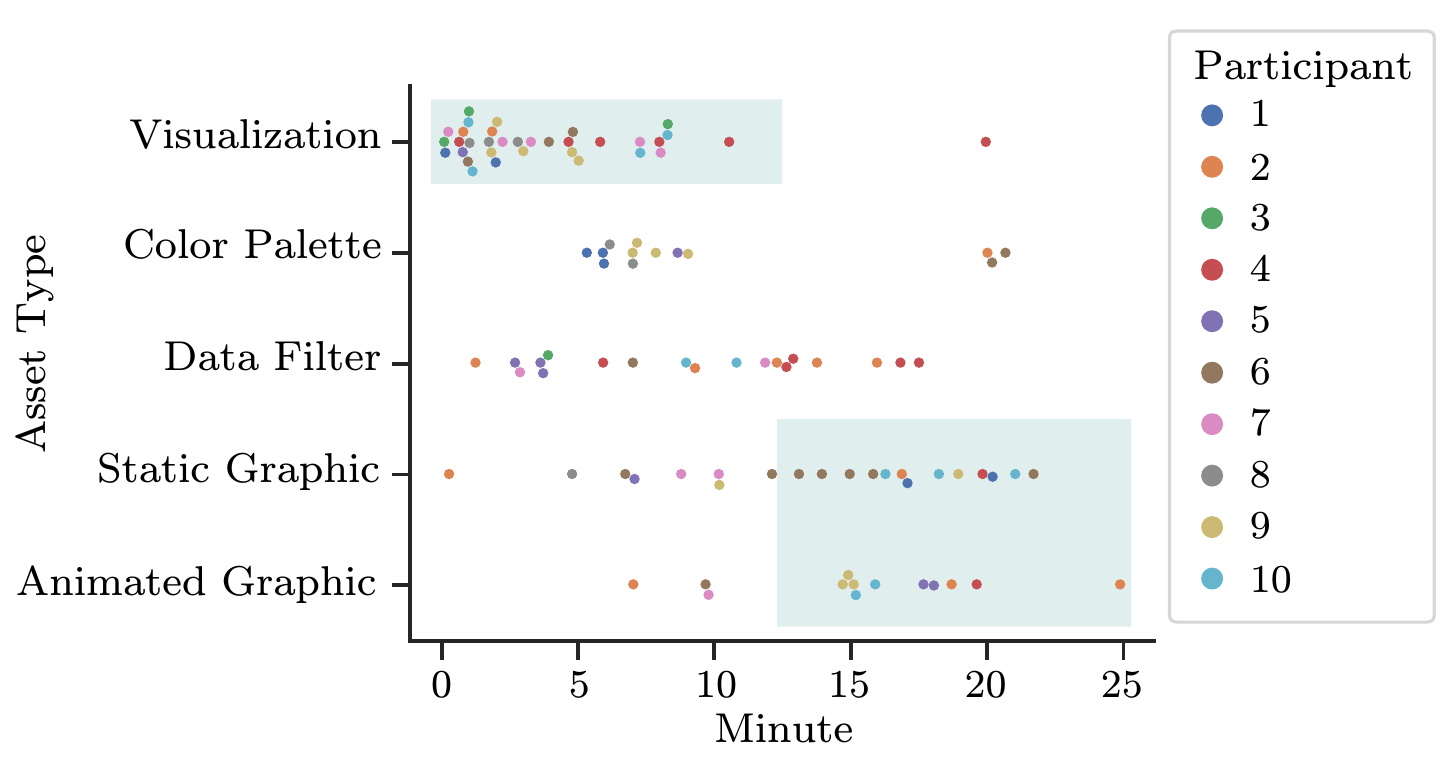}
  \caption{A mapping of asset types that participants initiated recommendations for over time. \chirevision{
  Each dot is an instance where a participant requested a specific asset type using the system. Each light green rectangle spans half of the allotted time, 12.5 minutes, and is used to visually highlight when a majority of recommendations for that asset type occurred. 
  } Most participants focused on generating visualizations during the first half (\chirevision{$50\%$ of asset interactions}), and graphics in the later half (\chirevision{$70\%$ of asset interactions}).}
  \label{fig:asset_type_over_minute}
  \Description{A scatter of what asset types were recommended at each minute. Points are color coded by the participant. There are shaded regions behind the points from minute 0-12.5 for visualization and from minute 12.5-25 for static and animated graphics.}
\end{figure}

\begin{figure*}
  \includegraphics[width=\textwidth]{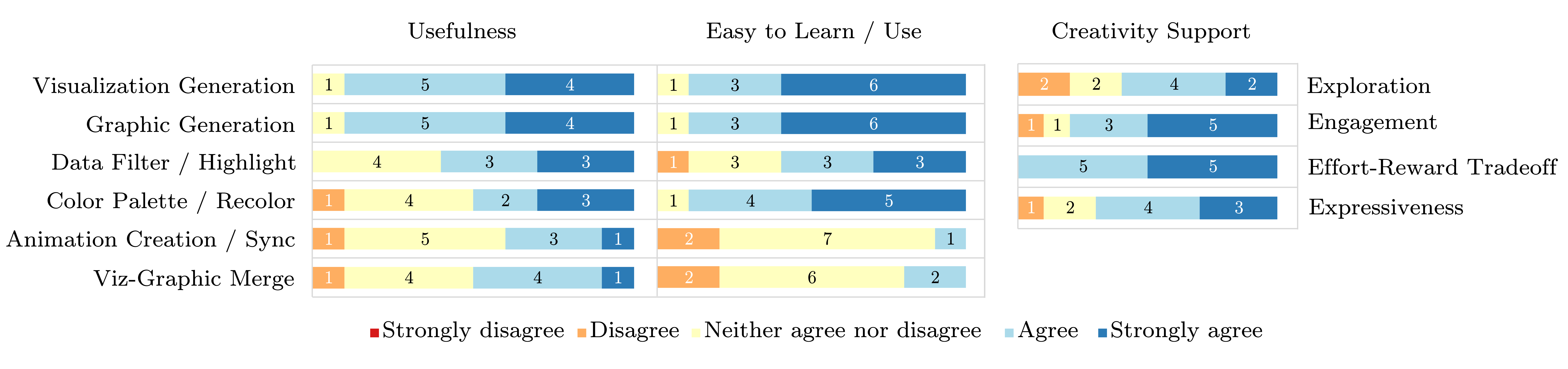}
  \caption{System usability scores for individual features and creativity support index scores obtained from 10 participants. Participants found the visualization generation and graphic generation features to be the most useful. Visualization generation, graphic generation, and color palette / recolor were the easiest to learn and use. Overall, all participants agreed that the end result was worth the effort.}
  \label{fig:usability_and_creativity}
  \Description{Stacked bar charts showing the 5 point Likert scale ratings participants made for creativity support as well as the usefulness and easiness to learn / use for each asset type.}
\end{figure*}

\subsubsection{Post-Survey \& Interview (\chirevision{25} minutes)}

At the conclusion of the tasks, participants filled out a post-survey about the usability of the individual features based on a subset of the SUS scale~\cite{brooke1996sus}, perceived similarity between their sketch and final infographic, overall satisfaction, and feelings of creativity support based on the Creativity Support Index~\cite{cherry2014quantifying}. They also participated in \chirevision{two} semi-structured interviews. \chirevision{The first} probed their thoughts on their authoring process, impressions of the system compared to other ones for infographic creation, feelings on the level of automation, use cases for the system, and the natural-language centered workflow. \chirevision{The second asked them to retroactively reflect on their asset generation choices with respect to their sketch and how their mental models evolved from beginning to end.}

\section{Results} \label{sec:user_study}


\subsection{Personas Induced by Workflow Patterns}

The distribution of how participants used mouse clicks are depicted in Figure~\ref{fig:clicks_per_participant}. While participants ranged quite widely in the number of total clicks they made ($\mu=242.7, \sigma=56.5$), all of them spent less than a quarter of these clicks on generating assets ($\mu=13.2\%, \sigma=5.7\%$), and \chirevision{the rest on arranging assets on the canvas}. One participant spent as little as 9 clicks total ($4.7\%$) producing the visual components they desired and devoted the rest of their interactions fine-tuning the layout, sizing, and captions. The comparatively fewer clicks required to obtain assets surprised some participants, who compared this to the tediousness of their prior workflows having to collect assets from different sources, which \textit{``really impedes creativity because when you're doing something creative, you get into the zone, but then it's like, oh wait, I need a cloud and then you spend the next 5--10 minutes elsewhere trying to find the cloud''} (P9). The mental workload of crafting the infographic is shifted to the design instead. \chirevision{From the asset generation clicks, we also broke down which types of} assets participants wanted over time, depicted in Figure \ref{fig:asset_type_over_minute}. \chirevision{Many participants spent the majority of their clicks on visualizations ($50\%$) in the first 12.5 minutes and wanted static and animated graphics ($70\%$) in the latter half}. In contrast, there were two small clusters where color palettes were more desirable--namely between the 5 to 10 minute mark immediately after visualizations and after 20 minutes after most of the other visual elements were added. This makes sense in context, as participants would want to change the color scheme of assets after they have been imported. These patterns align with those that can be found in traditional editing experiences, indicating the presence of \textit{legacy bias}~\cite{morris2014reducing} and indicating no steeper learning curves are introduced from disrupting old editing orders.

\chirevision{From our interviews, we also summarized the participants' verbalized mental models on these interaction patterns into two personas: 1) ``confident users'' who, either from their prior sketch or background knowledge, knew exactly what they wanted to depict and 2) ``exploratory users'' who did not and used the recommendations for ideation. The former group approached the workflow from a functional perspective, stating ``\textit{I tackled the largest and most important element first, which would be the visualization} (P7),'' while the latter felt that there were more options to explore for visualizations and the graphics would be dependent on them. The ``confident'' group also devoted their time into looking for the exact assets that matched their sketch; some succeeded easily (P5, P7), while others had to re-calibrate and find alternative options (P9, P10). Conversely, the ``exploratory'' group used the key message more as a guideline for their exploration, specifically using the message to group assets into visual sets. For example, visualizations and graphics relating to the same sentence in the message would be arranged spatially together on the canvas.}

\begin{figure*}
  \includegraphics[width=\textwidth]{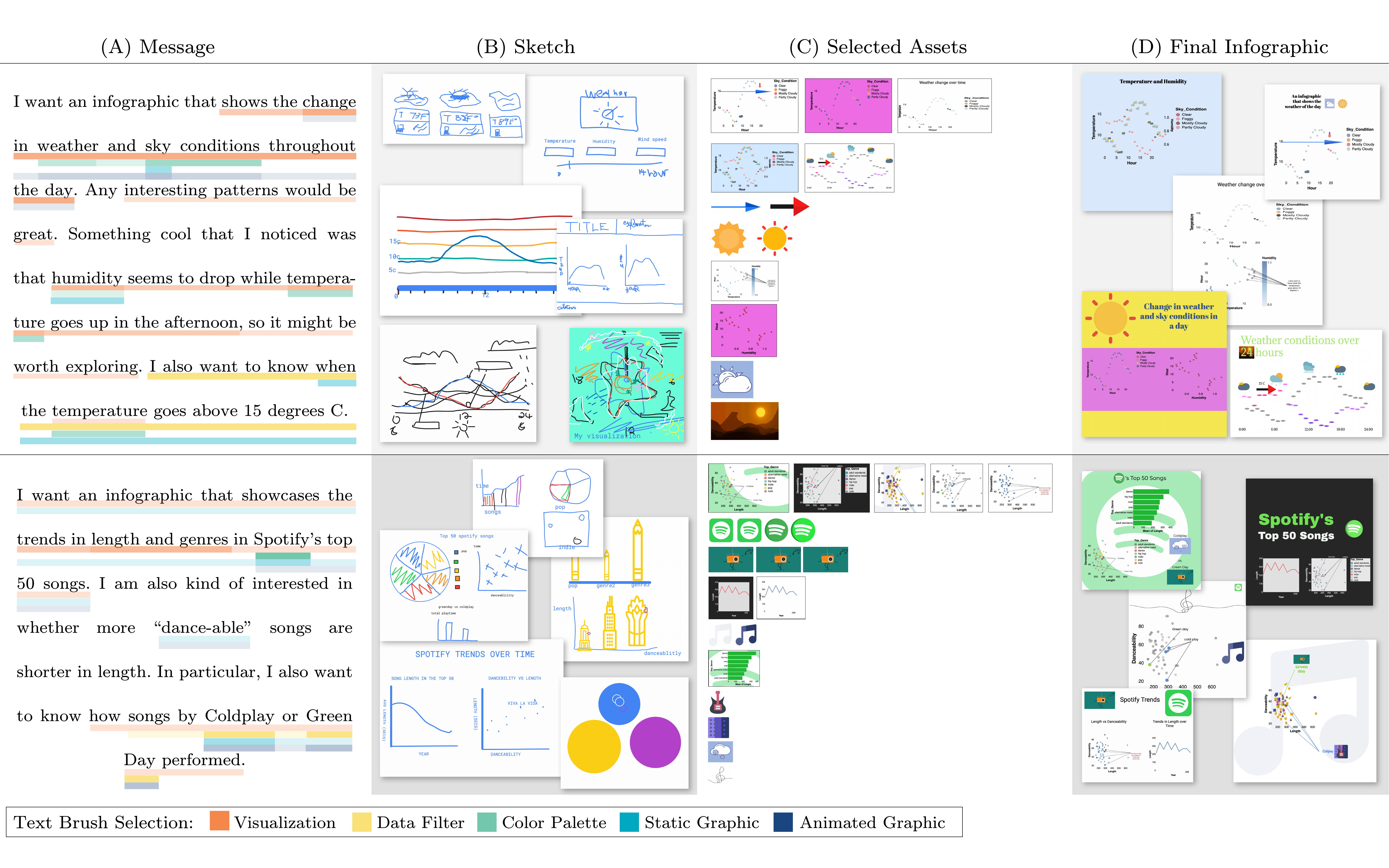}
  \caption{Overview of the participant authoring process. (A) A heat map overlaid on the messages provided indicates that there are trends in how users brush over text for different types of asset recommendations. (B) Sketches from the participants before using our system show divergent content and layouts. (C) Participants selected similar assets for their infographic. (D) The final infographic shows convergent content but divergent layouts.}
  \label{fig:participants_authoring_process}
  \Description{A 4x2 table. The 4 columns show the key message, the participant sketch, selected assets, and the final infographic. The message column contains the message itself and heatmap underlines showing what the participants brushed over, color coded by asset types. The sketch column contains different participant sketches evenly spaced out. The selected assets column contains individual assets pulled from the final infographics arranged by similar type. The final infographic shows the final infographics even spaced out. The 2 rows show each different message provided to the participants with the corresponding sketches, selected assets, and final infographics.}
\end{figure*}

\subsection{Effectiveness of Asset Recommendations and Interactions}

A summary of the usability scores for each system feature and overall feelings of creativity can be found in Figure \ref{fig:usability_and_creativity}. Participants agreed that the visualization and graphic recommendations are the most useful and easiest to use because they are the core components that make up the infographic. They found the data filter and color palette recommendations were comparatively less useful because the former is circumstantial and the latter only improves aesthetics and style, which reinforces the message but is not necessarily central to the message. Animation creation/sync and visualization-graphic merge functions are similarly useful, but harder to learn and use. Their use cases are more niche. Animation requires that the specific dataset has a time-oriented column to be animated over and visualization-graphic merge requires a visualization that has a legend with labels replaceable by visual substitutions. However, in the situations where they can be used, participants recognized that they could make the infographic more engaging. For example, P9 initially felt like they were \textit{``trying to find a use for animation,''} but upon more introspection about the dataset and the message, they \textit{``could imagine I could do something with animation sync. Like moving through the 24 hours of the GIF and matching that to the day night cycle of the visualization. And I feel like if I had the time to do that, that would actually be really cool and really unique.''}

Both the novices and experts agreed that the system allowed them to create designs without tedious interactions, was engaging, and allowed them to be expressive. All participants (5 agree, 5 strongly agree) felt that the resulting design they were able to produce was worth the effort it took to produce it. Comparing this workflow against existing workflows they would have used to create infographics, all participants stated that this system was faster. Specifically, P8 said, \textit{``Usually, I would probably use either Figma or Photoshop to lay down the layout. But I would have to leave the program to find graphics or go to Illustrator to make my own graphics. Here, I feel like I can make one that's decent without having to leave the program.''} In addition to the convenience of a centralized tool, the automatic recommendation of assets and suggestions to integrate them removes some of the barrier of \textit{``data science knowledge''} (P5) it takes to manually decide which bits of information to prioritize. Instead of the designer tunnel-visioning because \textit{``you have to have something very specific in mind before you make it (P4)''} with existing workflows, Epigraphics allowed participants to think about infographic composition more holistically (P9).

\begin{figure*}[t]
  \includegraphics[width=0.45\textwidth]{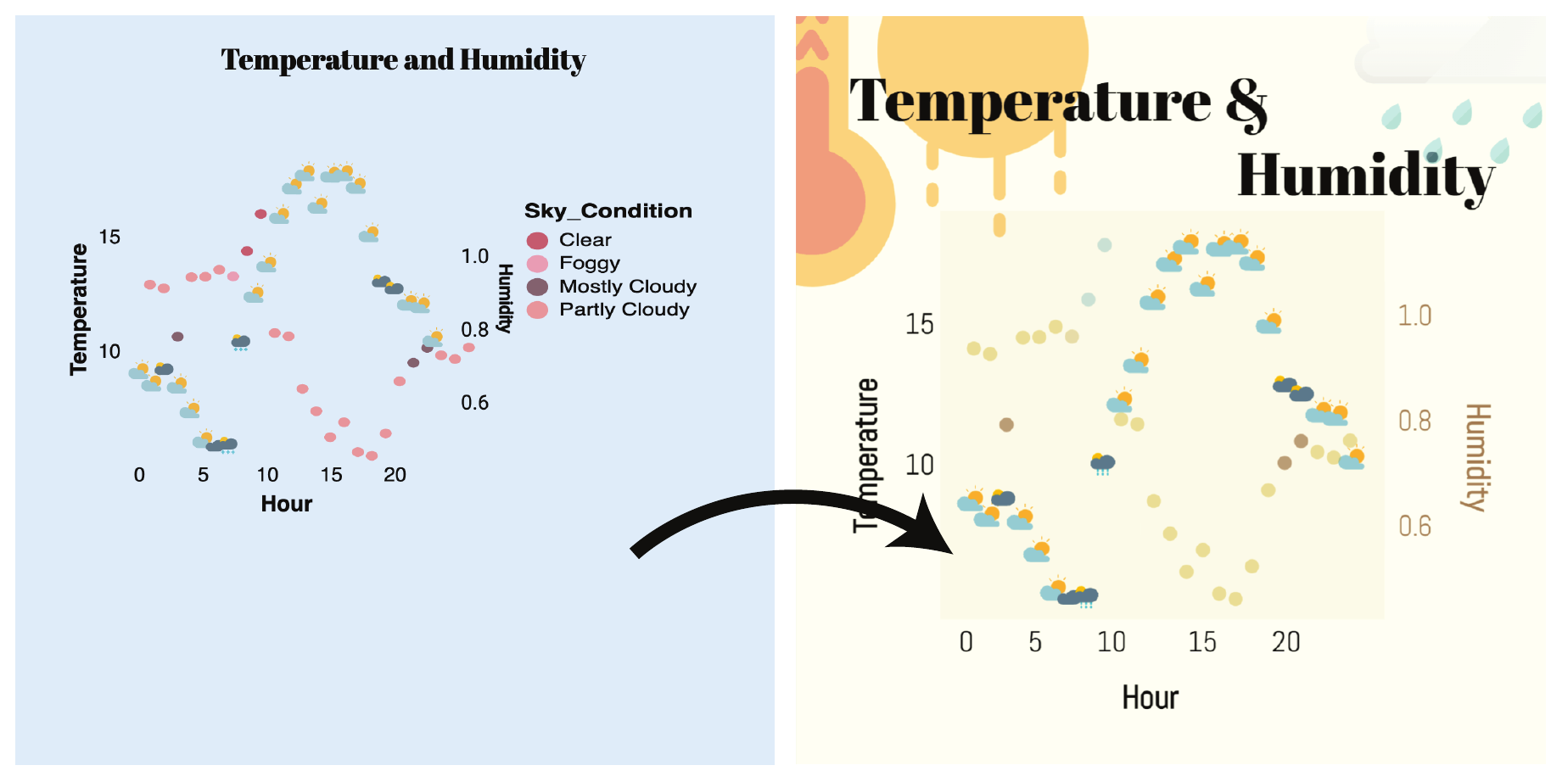}
  \includegraphics[width=0.45\textwidth]{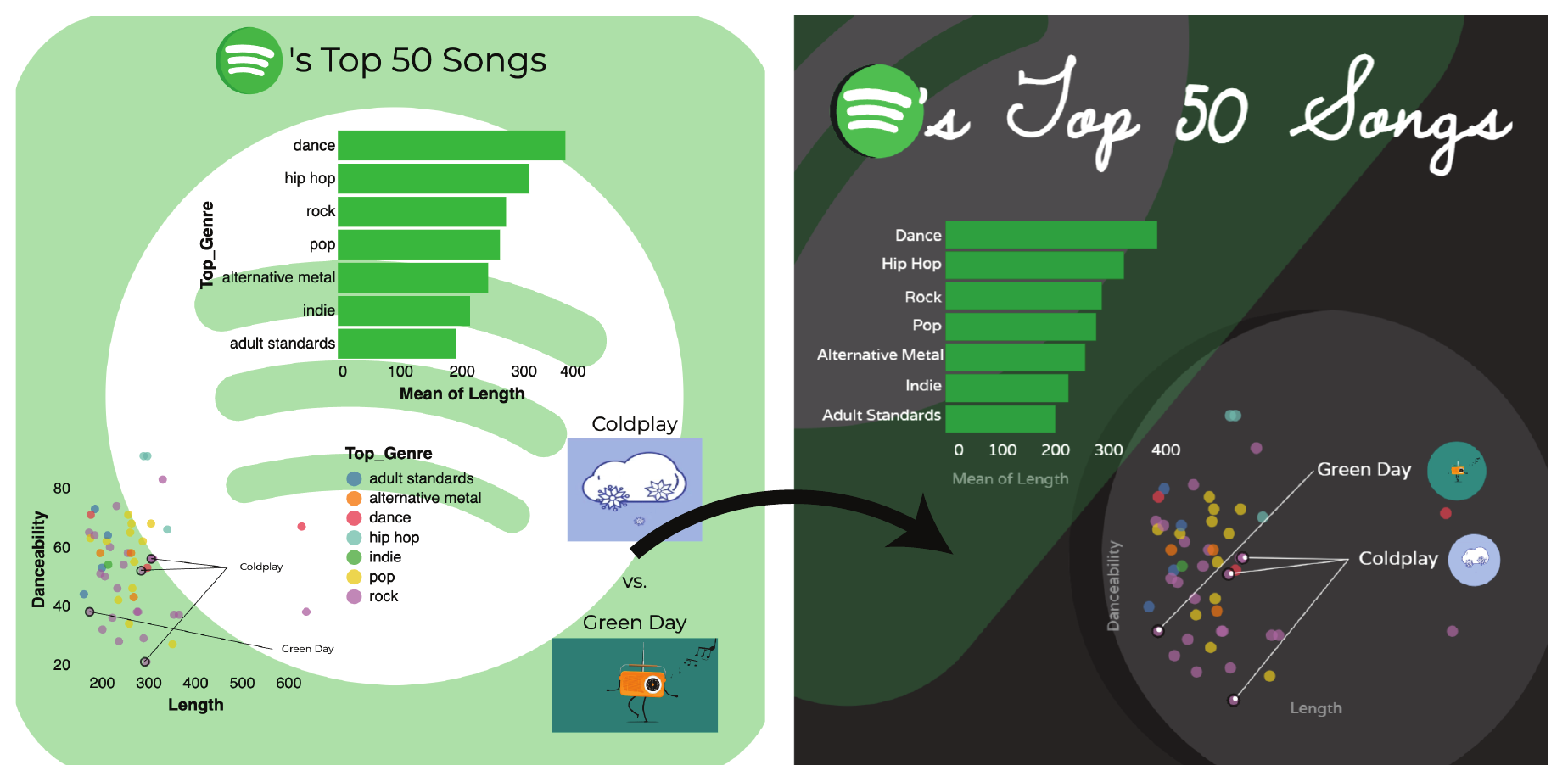}
  \vspace{-3mm}
  \caption{\chirevision{Before and after of two redesigns of participant (P1, P2) infographics by maintaining content and re-organizing layout. Modifying the background to enhance contrast, enlarging the title, rotating the logos, changing fonts, and grouping the plots more tightly can enhance visual cohesion.}}
  \label{fig:participants_infographic_redesign}
  \Description{Four infographics with arrows pointed from the first one to the second one and from the third one to the fourth one. The first infographic features two overlapping scatterplots on the top half and a small title. The second infographic contains the same overlapping scatterplots now enlarged to span the entire page, a larger title, and modified fonts. The third infographic features a small title on top, a green bar chart on the middle top, an annotated scatterplot on the bottom left, two images stacked on top of each other on the bottom right, and a background of the Spotify logo. The fourth infographic features a large title, the bar chart on the top left, the annotated scatterplot on the bottom right with the two images cropped as a circle next to each annotation, and a dark background with the Spotify logo rotated 45 degrees.}
  \vspace{-2mm}
\end{figure*}

\subsection{Comparing Fidelity to Message between Sketch and Infographic}

\chirevision{All participants felt that their final infographics aligned with the client-provided message. Figure \ref{fig:participants_authoring_process} depicts an overview of this message (A), their intermediate sketch (B), final infographic (D), and commonalities shared between the infographics (C) for the two datasets. Unprompted, the participants still displayed similar patterns in \textit{what} text they brushed for what types of assets. For example qualitative, more general statements such as ``change in weather and sky conditions throughout the day'' and ``trends in length and genres'' were used for visualizations, while specific sentence fragments such as ``temperature goes above 15 degrees C'' and ``how songs by Coldplay or Green Day performed'' were used for data filters. Graphics and color palettes concentrated on one or two keywords such as ``sky conditions,'' ``temperature,'' and ``Spotify'' for color palettes and ``sky,'' ``Coldplay,'' and ``Green Day'' for graphics. This resulted in similarities in the recommended and ultimately used assets in the infographic (Figure \ref{fig:participants_authoring_process}C). For the weather dataset, all participants used scatterplots of temperature versus hour or overlaid that on top of a humidity versus hour plot to depict the message ``throughout the day.'' To indicate specific temperatures, two participants shared the idea of using arrows as the symbol. Similarly, for the Spotify dataset, four participants added the Spotify logo and all of them appropriately highlighted data points referring to Coldplay and Green Day songs. In contrast, since the sketches (Figure \ref{fig:participants_authoring_process}B) were free-form, they displayed visually more disparate contents and layouts. In addition to more variety in chart types, each sketch also had different focuses; for example, some wanted to emphasize individual data points (P1, P3), others wanted to showcase general trends (P2, P5, P6, P8, P9), while the rest wanted to achieve a combination of both (P4, P7, P10).}

\chirevision{When asked to reflect upon the content differences, disregarding aesthetic polish, between their sketches and final infographics, eight of the ten preferred the infographic, and stated that the latter was more \textit{comprehensive} in presenting the information requested from the message. For example, P1 felt that in retrospect, their sketch did not sufficiently convey the humidity and temperature changes \textit{through the day} since they just drew three boxes of humidity and temperature at three specific time stamps. Similarly, two of the five Spotify sketches fail to mention Coldplay or Green Day at all, despite their emphasis in the key message. Multiple participants mentioned that the \textit{physical} action of brushing repeatedly over text encouraged them to ``\textit{cover all the bases}'' (P2), whereas this was less reinforced by just looking at the key message in the sketch task. For visualizations, since the brushed text is used under the hood as a query to reduce the number of columns into similar subsets for different users, all the participants ended up choosing from a ``standardized'' set of charts that shared similar axes and annotations. Similarly, graphics and color palettes were also standardized because the text-to-graphic and text-to-color provided unified representations for user intent and translated them to assets. For example, participants understood they had to reference Spotify somewhere in their infographic; two wrote the words ``Spotify'' on their sketch. This intent was physically translated to variations of the Spotify logo across all the infographics and reflected in the green/black color palettes used.
}

Overall, participants felt that using brushing over natural language to explore intent was a \textit{``cool [way] to automatically generate assets''} (P7). They pointed out that this was especially true for people \textit{``who are not good at math and statistics''} (P4) because they could verbally describe how they want the visualizations to combine with images, colors, or highlights, but may not necessarily have the prior knowledge to manually manipulate assets. However, some participants (P2, P5, P7, P9, P10, etc.) did want more extensive customization capabilities and alternate style recommendations for the assets after they were added to the canvas, such as deviating away from the flatness of the visualizations and static graphics to a \textit{``watercolour style''} (P9) or expanding the visualization to adopt more unconventional compositions (P1, P10). They expressed desires to achieve these results via text, as \textit{``I have to write everything down when I'm brainstorming, I write things. I don't draw things''} (P9).

\subsection{Reflection on Final Infographic Quality}
\chirevision{The final infographics created by the participants are located in Figure~\ref{fig:participants_authoring_process}D. In comparison to the example ones in the gallery of Figure \ref{fig:exemplar_gallery}, we note that they are less visually appealing. This disparity may be partially attributed to the differences in key message intent. For example, specifically comparing the user infographics for the Spotify dataset against Figure~\ref{fig:exemplar_gallery}E, which was constructed from the same dataset, we note that the key messages in the former consisted of more exploratory tasks. The participant key message wanted 1) trends in length and genres, 2) whether more ``dance-able'' songs are shorter, and 3) songs by Green Day or Coldplay, whereas the gallery key message wanted to showcase 1) indie songs have the highest energy and 2) hip hop songs are the most dance-able. This effect was intended as we wanted the participants to fully explore the system functionalities. As a side-effect, participants were more focused on using the recommendations to effectively identify trends or to answer the dance-ability question, and less focused on asset layout. However, the system was ultimately able to support users in adding the appropriate infographic components that addressed the message onto the canvas. From this point, we then argue that improving the appearance of the final infographic through asset layout once all the components are on the canvas does not take many extra steps. Figure~\ref{fig:participants_infographic_redesign} demonstrates how assets within two of the participant infographics can be re-arranged to generate a more visually effective infographic with the same content.}

\chirevision{But why didn't the participants perform these steps? While \systemname generates infographic components, it does not necessarily provide message-sourced visual groupings or font recommendations for components. Thus, to author more visually appealing final infographics, participants may require more background in layout design. Supporting this ``visual impressiveness'' is a trade-off between automation and freedom of interaction; while novices would rely on automation more to generate conventionally appealing designs, experts would prefer more flexibility of expression. Our work does not aim to replace designers' expertise, but rather to provide them with tools to extend their expertise. Future work could balance automation and interaction more during this final curation step to minimize the visual disparities of infographic outcomes.}









\section{Discussion}

Although the notion of a text-based workflow took some participants (P1, P6) time to adapt to, all users agreed that they would use this workflow in the future. P5 specifically noted that the system reminded them that \textit{``language allows users to do stuff that's uncommon and UI allows users to do stuff that's common.''} As \systemname is not necessarily the final form of text-powered authoring tools for data storytelling, we further summarize our findings about how natural-language sourced recommendations can support infographic content creation as design lessons \chirevision{and derive takeaways on core components of an effective key message below}.


\subsection{Text as a First-Class Object Effectively Standardizes Content and Mental Models}

When participants were provided the same text-based message, their output sketches were visually divergent. In contrast, when they used text brushing in \systemname, commonly brushing over the \textit{same} chunks, their final infographics were more visually cohesive. While the contents were not identical due to personal customization afterwards, the semantic information conveyed was similar. Participants overlapped in the chart \chirevision{axes used}, color families (more noticeable for the Spotify infographics), and images selected. This indicates that a `text as first-class object' paradigm has a standardizing effect over content authoring \chirevision{as keywords in the text 1) reduced dataset columns into semantically related subsets for visualizations and data filters and 2) attributed physical representations to \textit{implicit} intent for color palettes and graphics. Because the physical action of text brushing also incited users to critically think about and identify what words could be best used what asset types, it also provided some standardization over their mental models as they started to form mental mappings between text and asset type. From these interactions}, the assets added are also guaranteed to be relevant to the source message. This adherence to a \textit{focused} message implies that the resultant infographic is \chirevision{necessarily a \textit{comprehensive}} one \cite{dunlap2016getting, hernandez2021twelve, martin2019exploring, murray2017maximising}. 

One question that naturally arises then, is whether standardization restrains creativity. From the creativity support index scores in Figure \ref{fig:usability_and_creativity}, we see that most of the participants still agreed that the system allowed them to be very expressive. One participant specifically said, \textit{``Although it was a bit constraining, because there was such a variety with what the tool gave you within those constraints, I feel like it gave me things I would have never thought of.''} Other participants (P2, P5, P9), when reflecting on each type of asset generation during the post-survey, continued to talk unprompted about new designs they wanted to test by combining the assets they already had. The irony is that constraints within standardization made users \textit{more} creative because it encouraged them to think outside the box, and this resulted in more personally interesting outcomes. These sentiments align with prior studies that found that \textit{design constraints} \cite{stokes2001variability}, a set of boundaries set to a creative task, can stimulate creativity as opposed to suppress it \cite{caniels2015organizing, rietzschel2014effects}. Thus, we argue that text-sourced standardization of creative content is such a design constraint that can be incorporated into interactive tools beyond those for infographic authoring and can have a positive effect on practiced creativity. 


\subsection{Message-based Content Recommendations Empower Big Picture Thinking}

Despite the message-based approach being conducive to standardizing content, the layouts of the final participant infographics remained varied. \chirevision{Although some (P2) grouped assets based on similar semantics of the message they were sourced from, how this grouping occurred was unstructured.} For example, participants may place one visualization in the center of the page and surround it with images or stack two visualizations side-by-side either horizontally or vertically with the title either on the top left, top middle, or top right. Some used the recommended highlight functionality to emphasize data points while others manually added an arrow \chirevision{symbol} to achieve the same effect. Depending on how they were placed, the same graphics served different functions: as backgrounds, decorative accents for the title, attention drawers to highlighted data points, etc. The amount of deliberation in these different decisions is reflected in the large proportion of clicks during the authoring task that was devoted to canvas interactions. 
But what does flexibility in layout, but constraint in content mean? According to P9, this dichotomy indicates, \textit{``I was more focused on design. I think it meshes very naturally into the creative process because you're just focused on the big picture. Things like composition, like symbolism in the visuals.''} Attention is instead shifted away from the specificity of individual components to the design as a holistic view. \chirevision{This helped some participants avoid tunnel visioning, which occurred during the sketching task, and helped others re-calibrate their expectations of how they wanted to present the message.}

While systems that can effectively automate layout exist \cite{schrier2008adaptive, tabata2019automatic, guo2021vinci}, we argue that maintaining complete user autonomy over design layout has benefits, especially if the asset generation process is already automated. Particularly for personalized designs, active thinking about design alternatives allows users to avoid a linear design process \cite{swearngin2020scout} and create non-derivative motifs. More importantly, combining the standardization of assets with flexibility in layout supports \textit{concurrent} convergent and divergent thinking, both of which necessarily occurs ``cycling repeatedly'' \cite{vidal2010creative} in the creative process \cite{goldschmidt2016linkographic, perkins1992topography, fricke1996successful}. Perkins \cite{perkins1992topography} makes this even more explicit, stating that \textit{``inventive people are mode shifters''} between convergent and divergent thinking; tools that incorporate \textit{both} into the workflow can thus more effectively facilitate innovation.


\subsection{Accelerating Asset Generation Facilitates Rapid Iteration across Mental Models}


\chirevision{We previously identified two personas for patterns of asset generation: 1) the ``confident'' user who looks for specific assets that match their preconceived beliefs and 2) the ``exploratory'' user who generates assets to understand and brainstorm what final component they want. While the divide was not completely clear-cut with two exceptions, most of our novice-leaning users were ``exploratory,'' while the expert-leaning users were ``confident.'' In both instances, the message-based approach} offloaded click counts from the asset generation process so users focused more on canvas interactions. \chirevision{This means that the ``confident'' user had more time and space to either recreate their visions or re-align them with either the message or tool capabilities. Conversely for the ``exploratory'' user, they were able to see a plethora of feasible visual stimuli to explore potential representations of what they want to convey from the inventory.} P4 reinforces this, stating, \textit{``It would be really useful for not even sketching or brainstorming but like pre-brainstorming. I feel that usually this is hard, but it can give you just a wide scattershot of random graphs, images, or points to emphasize. And it's that relationship in this system that is actually really interesting}.'' \chirevision{The centralized authoring environment and reduction of search space due to the querying nature from message based interactions allowed both groups to iterate through ideas and assets quickly.} 

It is often difficult to design a system for both novices and experts due to the differences in how they complete a specific task within a specific tool. \chirevision{The novice is often} ``depth-first'' and \chirevision{considers many} sub-solutions in depth before making decisions, whereas experts can \chirevision{skip that step by} mentally removing themselves from specific examples to \chirevision{envision} more general or abstract concepts \cite{cross2004expertise}. We found that the few-clicks paradigm for assets afforded by messages empowers both, supporting exploration more akin to pre-brainstorming or prototyping for new users and rapid assembly of complete infographics for experienced ones. Both types of users are thus able to rapidly iterate on their personal goals. Similarly, other existing systems that intentionally minimize click fatigue for performance enhancements have reported benefits such as reducing frustration and increasing engagement \cite{bao2006fewer}, both of which support the rapid iteration process. The novice can thus quickly and enjoyably gain greater familiarity with both visualization, design concepts, and the system itself \chirevision{until they become an expert and can rapidly create} complete infographics of their own. 


\subsection{\textbf{Strategies for an Effective Key Message and How It Might Fail}}
\chirevision{
Based on our case studies and user study, we further reflect on the core properties of our message-driven approach, the message itself. Given the assets \systemname can recommend, an effective key message can be deconstructed into constituents of keywords or phrases, yet should also be a semantically correct, stand-alone summary of the envisioned infographic. Although there are no restrictions on what a user could brush over for each asset type, certain phrases can lead to greater success in recommendation quality. Consider the key message used in the first case study (Section \ref{sec:case_study_one}), ``\textit{A canary flapping its wings based on traced body positions taken from slow-motion video captures of the bird, highlighting its upstrokes}.'' Visually descriptive noun phrases such as ``canary,'' ``canary flapping its wings,'' ``video captures,'' and ``bird'' can either allude to potential sources for static and animated graphics or for thematic color families. Qualitative or quantitative verb phrases that modify these noun phrases such as ``flapping its wings based on traced body positions'' or ``taken from slow-motion video captures'' are good candidates for visualizations. Specific adjective prepositional descriptors that modify the verb phrases such as ``highlighting its upstrokes'' can be used for data filters. Thus, depending on what components the user wishes to include, a standard key message could consist of a mixture of 1) noun phrases, 2) verb phrases, and 3) adjective prepositional descriptors.
}

\chirevision{
However, undesired assets may be generated when message phrases are brushed over incorrectly. Some obvious failure cases from the user study can be seen in Figure~\ref{fig:participants_authoring_process}A, where the brushed text provided either too much or too little information. For example, some users brushed the whole first sentence in the Spotify message, which consisted of multiple phrases that could allude to different asset types, to request a graphic. The subsequent recommendations were too ambiguous and varied to be useful. Conversely, another user mistakenly brushed the word ``conditions'' to request a color palette, which was too little context and led to no recommendations. Thus, while the key message itself should be comprehensive, which words or phrases are selected also require precision and thought. In such a heavy text-dependent workflow, we therefore emphasize the importance of the brushing interaction; since trial-and-error is required in iterating through desired assets, streamling this interaction could increase efficiency. 
}

\subsection{Limitations and Future Work}

Participants reported that the main limitation of the system was that it didn't have customization capabilities comparable to other design software like Illustrator or Figma. They also wanted to create stylized visualizations in painterly or sketch-like fashions beyond the flat vector appearance. The system also did not support the entire breadth of visualization chart types. Thus, future work could expand on the expressiveness of the graph styles via application of SVG filters \cite{zhou2023filtered} and the addition of more graph types. Furthermore, the current retrieval process for graphic generation could be substituted with high-fidelity text-to-image and text-to-animation, instead, which will further expand the variety of infographics that can be rapidly created. Since it is modular, the final form of \systemname could also easily serve as a plugin for existing design tools. A thorough comparison with other semi-automatic systems for infographic authoring tools can also reveal more nuances about the trade-offs of a message-sourced approach. 
\section{Conclusion}

Graphical authoring tools have historically started with the canvas, prescribing a workflow that focuses on visuals first. By applying generative models to translate a message into components of an infographic, \systemname explores an approach to rapidly prototyping infographics starting with the author's intent. This workflow automatically generates data visualizations, graphics, colors, highlights, and animations to help designers assemble a complete infographic that \chirevision{conveys a cohesive theme}. Participants noticed the integrated workflow allowed them to switch back and forth between text and canvas within the same application as they combined assets together while still thinking about the core message. \chirevision{It induced greater infographic fidelity to the message through the physical action of brushing, as well as} affected what was ultimately produced. \chirevision{It also led} to a convergence of \chirevision{both content and user mental models, while maintaining} a diversity of styles in the final infographics, \chirevision{and helped users ideate more holistically despite varying levels of expertise. Both text and canvas have their advantages--the text can support semantically nuances in its input and the canvas precision in its output;} while text is linear, the canvas is multidimensional in layout and theme. \chirevision{By combining the two together}, \systemname adds that second dimension to infographics authoring that empowers rapid iterations \chirevision{to produce a first draft that comprehensively conveys a} coherent visual message.

\begin{acks}
We would like to thank Eunyee Koh, Haijun Xia, Ji Won Chung, and the AEL group at Adobe Research for their feedback that helped shape this manuscript.
\end{acks}

\bibliographystyle{ACM-Reference-Format}
\bibliography{sample-base}

\end{document}